\documentclass[aps,tightlines]{revtex4}%
\usepackage{amsfonts}
\usepackage{amsmath}
\usepackage{amssymb}
\usepackage{graphicx}%
\begin{document}
\title{Monogamy constraints on entanglement of four-qubit pure states}
\author{S. Shelly Sharma\thanks{Retired} }
\email{shelly@uel.br}
\affiliation{Departamento de F\'{\i}sica, Universidade Estadual de Londrina, Londrina
86051-990, PR Brazil}
\author{N. K. Sharma\thanks{Retired} }
\email{nsharma@uel.br}
\affiliation{Departamento de Matematica, Universidade Estadual de Londrina, Londrina
86051-990, PR Brazil }

\begin{abstract}
We report a set of monogamy constraints on one-tangle, two-tangles,
three-tangles and four-way correlations of a general four-qubit pure state. It
is found that given a two-qubit marginal state $\rho$ of a four qubit pure
state $\left\vert \Psi_{4}\right\rangle $, the non-Hermitian matrix
$\rho\widetilde{\rho}$ where $\widetilde{\rho}$ $=\left(  \sigma_{y}%
\otimes\sigma_{y}\right)  \rho^{\ast}\left(  \sigma_{y}\otimes\sigma
_{y}\right)  $, contains information not only about the entanglement
properties of the two-qubits in state $\rho$ but also about three tangles
involving the selected pair as well as four-way correlations of the pair of
qubits in $\left\vert \Psi_{4}\right\rangle $. To extract information about
tangles of a four-qubit state $\left\vert \Psi_{4}\right\rangle $, the
coefficients in the characteristic polynomial of matrix $\rho\widetilde{\rho}$
are analytically expressed in terms of $2\times2$ matrices of state
coefficients. Four-tangles distinguish between different types of entangled
four-qubit pure states.

\end{abstract}
\maketitle

\section{Introduction}

Entanglement is not only a necessary ingredient for processing quantum
information \cite{niel11} but also has important applications in other areas
such as quantum field theory \cite{cala12}, statistical physics \cite{sahl15},
and quantum biology \cite{lamb13}. Multipartite entanglement is a resource for
multiuser quantum information tasks. Bipartite entanglement is well understood
as there is concise result to entanglement classification problem. For
bipartite systems, the notion of maximally entangled states is independent of
the specific quantification of entanglement. However, since the mathematical
structure of multipartite states is much more complex than that of bipartite
states, the characterization of multipartite entanglement is a far more
challenging task \cite{horo09}. Even the identification of maximally entangled
states in multi-party systems is highly non-trivial.

Walter et al. \cite{walt13} used an algebraic geometry approach to show that
single particle states are a rich source of information on multiparticle
entanglement. In a recent letter \cite{arxiv2021}, we have shown that
two-qubit subsystems of an N-qubit state contain information about the
correlations beyond two-qubit entanglement. In this context, four-qubit pure
states offer an interesting case study, since a good part of residual
correlations can be identified as contributions from seven four-tangles as
defined in this article. One of the key features of multipartite correlations
that separates quantum-ness from classicality is monogamy of entanglement.
Monogamy of quantum entanglement refers to shareability of entanglement in a
composite quantum system. Monogamy relation for entanglement of three-qubit
states, known as CKW inequality, was reported in a seminal paper by Coffman,
Kundu, and Wootters \cite{coff00}. Recent efforts to find monogamy relations
satisfied by tangles of four-qubit states include refs.
\cite{regu14,regu15,regu16,chri18,shar18,daekil18}. A detailed analysis of
residual correlations for four-qubit pure states, yields monogamy constraints,
reported in this article.

One-tangle is known to quantify the entanglement of a single qubit with the
rest of the composite system in an $N$-qubit pure state, whereas two-tangle
(or concurrence as defined in ref. \cite{hill97,woot98} ) is a measure of
entanglement of two qubits. One tangle, defined as $\tau_{1|2...N}=2\left(
1-\text{Tr}\left(  \rho^{2}\right)  \right)  $, quantifies essentially the
mixedness of single-qubit marginal state $\rho$. Likewise, mixedness of a
two-qubit marginal state of an $N-$qubit pure state is due to $2-$way,
$3-$way, ....,$N-$way correlations of two qubits with $N-2$ qubits. In
\cite{arxiv2021}, it has been shown, analytically, that two-tangle can be
written as the difference of two terms, where the first term contributes to
one-tangle while the second term is a function of degree $8$, $12$ and $16$
local unitary invariant functions of state coefficients. Since two-tangle and
the first term are calculable quantities, the difference gives quantitative
information about correlations beyond two-way correlations. Reported monogamy
constraints are functional relations satisfied by entanglement of a single
qubit to the rest of the system (one-tangle), the entanglement of two-qubit
marginal states (two-tangles), entanglement of three-qubit marginal states due
to three-way correlations (three-tangles), and the residual correlations
written as functions of four-qubit unitary invariant functions of state
coefficients (four-tangles). By identifying quantitatively the contributions
of three-tangles and four-tangles to correlations beyond two-tangles, it is
possible to know how entanglement is distributed in subsystems of the
four-qubit pure state. Monogamy of entanglement has potential applications in
areas of physics such as quantum key distribution \cite{terh04,pawl10,gisi02},
classification of quantum states \cite{dur00,gior11,prab12}, frustrated spin
systems \cite{ma11,rao13}, and even black-hole physics \cite{lloy14}.

Monogamy constraints are closely related to classification of entangled
states. Four-tangles, when used to label the entanglement classes of ref.
\cite{vers02} along with three-tangles and two-tangles unambiguously
distinguish between different types of entanglement due to four-way
correlations. In the case of four-qubit states, information from single
particle states \cite{walt13} already points to different entanglement types
in four-qubit pure states, however, information from two-qubit subsystems
quantifies the entanglement of the states due to four-way correlations in each class.

Two-tangle, one-tangle, three-tangle and necessary unitary invariants are
defined in Section II through Section VI, with our main results presented in
Sections VII, VIII and IX. Analysis of tangles of a four-qubit GHZ state and
Cluster state are in Sections X and XI. Monogamy of four-qubit correlations in
a special subset of four-qubit states $L_{a,ia,\left(  ia\right)  _{2}}$ is
discussed in section XII. Tangle based classification of four-qubit states is
discussed in section XIII. Section XIV on entanglement transfer using a simple
circuit model illustrates how two-way correlations of a pair of qubits leaks
into environment through successive interactions of one of the qubits of the
pair. Concluding remarks follow in section XV.

\section{Definition of two-tangle}

Two-tangle or concurrence, a well known measure of two-qubit entanglement
\cite{hill97,woot98} is an entanglement monotone. A generic two-qubit pure
state in computational basis reads as%
\begin{equation}
\left\vert \Psi_{12}\right\rangle =\sum_{i_{1},i_{2}}a_{i_{1}i_{2}}\left\vert
i_{1}i_{2}\right\rangle ;\quad(i_{m}=0,1),
\end{equation}
where $a_{i_{1}i_{2}}$ are the state coefficients. The indices $i_{1}$ and
$i_{2}$ refer to the state of qubits $A_{1}$ and $A_{2}$, respectively.
Entanglement of qubit $A_{1}$ with $A_{2}$ is quantified by two-tangle defined
as%
\begin{equation}
\tau_{1|2}\left(  \left\vert \Psi_{12}\right\rangle \right)  =2\left\vert
a_{00}a_{11}-a_{10}a_{01}\right\vert .
\end{equation}
Consider the action of a unitary transformation $U^{j}=\frac{1}{\sqrt
{1+\left\vert x\right\vert ^{2}}}\left[
\begin{array}
[c]{cc}%
1 & -x^{\ast}\\
x & 1
\end{array}
\right]  $ on qubit $A_{j}$. We can verify that%
\begin{equation}
U^{1}\tau_{1|2}\left(  \left\vert \Psi_{12}\right\rangle \right)  =U^{2}%
\tau_{1|2}\left(  \left\vert \Psi_{12}\right\rangle \right)  =\tau
_{1|2}\left(  \left\vert \Psi_{12}\right\rangle \right)  .
\end{equation}

Two-tangle of a mixed state $\rho=\sum\limits_{i}p_{i}\left\vert \phi
_{12}^{\left(  i\right)  }\right\rangle \left\langle \phi_{12}^{\left(
i\right)  }\right\vert $ is constructed through convex roof extension as%
\begin{equation}
\tau_{1|2}\left(  \rho\right)  =2\min_{\left\{  p_{i},\phi_{12}^{\left(
i\right)  }\right\}  }\sum\limits_{i}p_{i}\left\vert a_{00}^{\left(  i\right)
}a_{11}^{\left(  i\right)  }-a_{10}^{\left(  i\right)  }a_{01}^{\left(
i\right)  }\right\vert . \label{mix2tangle}%
\end{equation}
Specifically, two-tangle \cite{hill97,woot98} of a two-qubit state $\rho_{12}$
is given by%
\begin{equation}
\tau_{1|2}\left(  \rho\right)  =\max\left(  0,\sqrt{\lambda_{1}}-\sqrt
{\lambda_{2}}-\sqrt{\lambda_{3}}-\sqrt{\lambda_{4}}\right)  , \label{2tangle}%
\end{equation}
where $\lambda_{1}\geq\lambda_{2}\geq\lambda_{3}\geq\lambda_{4}$ are the
eigenvalues of non-Hermitian matrix $\rho\widetilde{\rho}$ with $\widetilde
{\rho}$ $=\left(  \sigma_{y}\otimes\sigma_{y}\right)  \rho^{\ast}\left(
\sigma_{y}\otimes\sigma_{y}\right)  $. Here $\ast$ denotes complex conjugation
in the standard basis and $\sigma_{y}$ is the Pauli matrix. In the most
general case, the characteristic polynomial of $\rho\widetilde{\rho}$ has the
form%
\begin{equation}
x^{4}-x^{3}n_{4}+x^{2}n_{8}-xn_{12}+n_{16}=0 \label{poly}%
\end{equation}
where the coefficients $n_{d}$ are given by%
\begin{equation}
n_{4}=tr\left(  \rho\widetilde{\rho}\right)  ;n_{8}\left(  \rho\right)
=\frac{1}{2}\left(  \left(  tr\rho\widetilde{\rho}\right)  ^{2}-tr\left(
\rho\widetilde{\rho}\right)  ^{2}\right)  ; \label{n4n8}%
\end{equation}%
\begin{equation}
n_{12}=\frac{1}{6}\left(  \left(  tr\rho\widetilde{\rho}\right)
^{3}-3tr\left(  \rho\widetilde{\rho}\right)  tr\left(  \rho\widetilde{\rho
}\right)  ^{2}+2tr\left(  \rho\widetilde{\rho}\right)  ^{3}\right)  ;
\label{n12}%
\end{equation}%
\begin{equation}
n_{16}=\det\left(  \rho\widetilde{\rho}\right)  . \label{n16}%
\end{equation}
Matrix elements of a two-qubit mixed state $\rho$ are degree-two functions of
state coefficients of the pure state from which $\rho$ has been obtained. As
such a given coefficient $n_{d}\left(  \rho\right)  $ is a unitary invariant
function of state coefficients of the pure state of which $\rho$ is a part.
The subscript $d$ refers to the degree of the invariant. Defining $C\left(
\rho\right)  =\sqrt{\lambda_{1}}-\sqrt{\lambda_{2}}-\sqrt{\lambda_{3}}%
-\sqrt{\lambda_{4}}$, we can verify that for $C\left(  \rho\right)
=\pm\left\vert C\left(  \rho\right)  \right\vert $ the coefficient $n_{4}$
satisfies the relation%
\begin{equation}
n_{4}=\left\vert C\left(  \rho\right)  \right\vert ^{2}+\sqrt{4n_{8}%
+8\sqrt{n_{16}}\pm8\sqrt{f_{16}}}, \label{n4rho}%
\end{equation}
where $f_{16}\geq0$ is defined as%
\begin{equation}
f_{16}=\sqrt{n_{16}}\left\vert C\left(  \rho\right)  \right\vert ^{2}\left(
n_{4}-\left\vert C\left(  \rho\right)  \right\vert ^{2}\right)  +n_{12}%
\left\vert C\left(  \rho\right)  \right\vert ^{2},, \label{f16}%
\end{equation}
To obtain Eq. (\ref{n4rho}) we used the expressions for the coefficients
$n_{d}$ ($d=4,8,12,16$) in terms of eigenvalues of matrix $\left(
\rho\widetilde{\rho}\right)  $ and the condition $\lambda_{1}\geq\lambda
_{2}\geq\lambda_{3}\geq\lambda_{4}$. Derivation of Eq. (\ref{n4rho}) is given
in Appendix \ref{A}. Since by definition $\tau_{1|2}\left(  \rho\right)
=\max\left(  0,C\left(  \rho\right)  \right)  $, we may rewrite Eq.
(\ref{n4rho}) as%
\begin{equation}
n_{4}-\tau_{1|2}^{2}\left(  \rho\right)  =\sqrt{4n_{8}+\chi_{12}^{\pm}},
\label{2tanrho}%
\end{equation}
where
\begin{equation}
\chi_{12}^{+}=8\sqrt{n_{16}}+8\sqrt{f_{16}}, \label{chiroplus}%
\end{equation}
and%
\begin{equation}
\chi_{12}^{-}=8\sqrt{n_{16}}-8\sqrt{f_{16}}+2n_{4}\left\vert C\left(
\rho\right)  \right\vert ^{2}-\left\vert C\left(  \rho\right)  \right\vert
^{4}. \label{chirominus}%
\end{equation}
This is an important relation between coefficients $n_{d}$ and two-tangle.

\section{Two-tangles and one-tangle of an N-qubit pure State}

Expressions for one-tangle as well as polynomial coefficients of degree four
and eight in terms of state coefficients of an N-qubit pure state are written
down, in this section. A general $N-$qubit pure state in computational basis
reads as%
\begin{equation}
\left\vert \Psi_{12...N}\right\rangle =\sum_{i_{1},i_{2},...,N}a_{i_{1}%
i_{2}...i_{N}}\left\vert i_{1}i_{2}...i_{N}\right\rangle ;\ i_{m}=0,1.
\label{psiN}%
\end{equation}
Here $a_{i_{1}i_{2}...i_{N}}$ are complex state coefficients and the indices
$i_{1}$, $i_{2},...,i_{N}$ refer to the state of qubits at locations $A_{1}$,
$A_{2}$,..., $A_{N}$, respectively. State of qubit pair $A_{1}A_{j}$ is
$\rho_{1j}=Tr_{2,...,j-1,j+1,...N}(\left\vert \Psi_{12...N}\right\rangle
\left\langle \Psi_{12...N}\right\vert )$ with matrix elements given by%
\begin{equation}
\left(  \rho_{1j}\right)  _{i_{1}i_{j}k_{1}k_{j}}=\sum_{I}a_{i_{1}i_{j}%
I}a_{k_{1}k_{j}I}^{\ast},
\end{equation}
where index $I=\left\{  i_{2}i_{3}...i_{j-1}i_{j+1}...i_{N}\right\}  $ with
associated value $I_{v}\equiv%
{\displaystyle\sum\limits_{\substack{m=2\\m\neq j}}^{N}}
2^{m-2}i_{m}$. Writing the characteristic polynomial for $\rho_{1j}%
\widetilde{\rho}_{1j}$ it is found that for qubit pair $A_{1}A_{j}$ in state
$\rho_{1j}$,%
\begin{equation}
n_{4}\left(  \rho_{1j}\right)  =2\sum_{I\leq J}\left\vert D_{1jIJ}%
+D_{1jJI}\right\vert ^{2}, \label{n4ro1j}%
\end{equation}
where
\begin{equation}
D_{1jIJ}=a_{0\left(  i_{j}=0\right)  I}a_{1\left(  i_{j}=1\right)
J}-a_{1\left(  i_{j}=0\right)  J}a_{0\left(  i_{j}=1\right)  I}. \label{d1jIJ}%
\end{equation}
A simplified notation $I<J$ \ is being used when $I_{v}<J_{v}$. The functions
$\left(  D_{1jIJ}+D_{1jJI}\right)  $ are invariant with respect to unitary
transformations on the focus qubit and qubit $j$ and depending on the value of
$I$ and $J$ represent a sum of determinants of $2-$way, $3-$way,..., $N-$way
matrices of dimension $2$. One-tangle defined as $\tau_{1|2...N}=4\det\left(
\rho_{1}\right)  $ where $\rho_{1}=Tr_{A_{2}...A_{N}}(\left\vert \Psi
_{12...N}\right\rangle \left\langle \Psi_{12...N}\right\vert )$ and $\left(
\rho_{1}\right)  _{i_{1}k_{1}}=\sum_{i_{j},I}a_{i_{1}i_{j}I}a_{k_{1}i_{j}%
I}^{\ast}$ quantifies the entanglement of qubit $A_{1}$ with rest of the
system. One can verify that%
\begin{equation}
\tau_{1|2...N}=4\sum_{j=2}^{N}\left(  \sum_{I}\left\vert D_{1jII}\right\vert
^{2}+\sum_{I<J}\left\vert D_{1jIJ}\right\vert ^{2}\right)  . \label{1tanN}%
\end{equation}
Comparing Eq. (\ref{n4ro1j}) and Eq. (\ref{1tanN}), we obtain%
\begin{equation}
\tau_{1|2...N}=\sum_{j=2}^{N}\left(  n_{4}\left(  \rho_{1j}\right)
-X_{1j}\right)  , \label{1tann4N}%
\end{equation}
where the quantity $X_{1j}$ defined as
\begin{equation}
X_{1j}=2\sum_{I<J}\left(  D_{1jIJ}D_{1jJI}^{\ast}+D_{1jIJ}^{\ast}%
D_{1jJI}\right)  , \label{coh1j}%
\end{equation}
represents coherences. The sum of coherences, $\sum_{j=2}^{N}X_{1j}$, turns
out to be zero for $N$-odd and is equal to sum of unitary invariants of degree
two for $N$-even.

The coefficient $n_{8}\left(  \rho_{1j}\right)  $, written in terms of state
coefficients reads as%
\begin{equation}
n_{8}\left(  \rho_{1j}\right)  =\sum_{I,J,K,L}\left\vert
\begin{array}
[c]{c}%
\left(  D_{1jIJ}+D_{1jJI}\right)  \left(  D_{1jKL}+D_{1jLK}\right) \\
-\left(  D_{1jIL}+D_{1jLI}\right)  \left(  D_{1jKJ}+D_{1jJK}\right)
\end{array}
\right\vert ^{2}. \label{n8ro1j}%
\end{equation}
While the coefficient $n_{4}\left(  \rho_{1j}\right)  $ is a sum of squares of
moduli of two-qubit invariants, the coefficient $n_{8}\left(  \rho
_{1j}\right)  $ is a sum of squares of three-qubit invariants.

\section{Tangles of a three-qubit pure State and monogamy of entanglement}

For a three-qubit system, the entanglement measures are known to satisfy CKW
inequality \cite{coff00}. In this section, we establish the relation between
coefficients $n_{d}\left(  \rho\right)  $ and entanglement measures of a
three-qubit pure state%
\begin{equation}
\left\vert \Psi_{123}\right\rangle =\sum\limits_{i_{1},i_{2},i_{3}}%
a_{i_{1}i_{2}i_{3}}\left\vert i_{1}i_{2}i_{3}\right\rangle ,\quad\left(
i_{m}=0,1\right)  . \label{3qstate}%
\end{equation}
Using the notation of ref. \cite{shar16}, the determinants of negativity fonts
for the state $\left\vert \Psi_{123}\right\rangle $ are defined as $D_{\left(
A_{3}\right)  _{i_{3}}}^{00}=a_{00i_{3}}a_{11i_{3}}-a_{10i_{3}}a_{01i_{3}}$
(two-way), $D_{\left(  A_{2}\right)  _{i_{2}}}^{00}=a_{0i_{2}0}a_{1i_{2}%
1}-a_{1i_{2}0}a_{0i_{2}1}$ (two-way), and $D^{00i_{3}}=a_{00i_{3}}%
a_{11i_{3}+1}-a_{10i_{3}}a_{01i_{3}+1}$ (three-way). This set of determinants
is the same as that obtained by substituting $\ j=2,3$\ and $I\equiv\left\{
0,1\right\}  $, in Eq. (\ref{d1jIJ}). For example, taking $A_{1}$ as focus
qubit $D_{12i_{3}i_{3}+1}=D_{13i_{2}i_{2}+1}=D^{00i_{3}}$. One-tangle, defined
as $\tau_{1|23}\left(  \left\vert \Psi_{123}\right\rangle \right)
=4\det\left(  \rho_{1}\right)  $ where $\rho_{1}=Tr_{A_{2}A_{3}}(\left\vert
\Psi_{123}\right\rangle \left\langle \Psi_{123}\right\vert )$, quantifies the
entanglement of qubit $A_{1}$ with qubits $A_{2}$ and $A_{3}$. Three tangle
\cite{coff00} of $\left\vert \Psi_{123}\right\rangle $ is equal to four times
the modulus of a unitary invariant polynomial of degree four that is%
\begin{equation}
\tau_{1|2|3}\left(  \left\vert \Psi_{123}\right\rangle \right)  =4\left\vert
I_{3,4}\left(  \left\vert \Psi_{123}\right\rangle \right)  \right\vert ,
\label{3tan}%
\end{equation}
where%
\begin{align}
I_{3,4}\left(  \left\vert \Psi_{123}\right\rangle \right)   &  =\left(
D^{000}+D^{001}\right)  ^{2}-4D_{\left(  A_{3}\right)  _{0}}^{00}D_{\left(
A_{3}\right)  _{1}}^{00}\nonumber\\
&  =\left(  D^{000}-D^{001}\right)  ^{2}-4D_{\left(  A_{2}\right)  _{0}}%
^{00}D_{\left(  A_{2}\right)  _{1}}^{00}. \label{3invariant}%
\end{align}
Three-tangle of the mixed state $\rho_{123}$ is defined as the average of pure
state three-tangles, minimized over all complex decompositions $\left\{
p_{i},\left\vert \phi_{123}^{\left(  i\right)  }\right\rangle \right\}  $ of
$\rho_{123}$ that is%
\begin{equation}
\tau_{1|2|3}\left(  \rho_{123}\right)  =\min_{\left\{  p_{i},\left\vert
\phi_{123}^{\left(  i\right)  }\right\rangle \right\}  }\sum\limits_{i}%
p_{i}\tau_{1|2|3}\left(  \left\vert \phi_{123}^{\left(  i\right)
}\right\rangle \right)  . \label{mix3tan}%
\end{equation}
Here $p_{i}$ is the probability of finding the normalized three-qubit state
$\left\vert \phi_{123}^{\left(  i\right)  }\right\rangle $ in the mixed state
$\rho_{123}$.

The relation between a matrix element of the state $\rho_{12}=Tr_{A_{3}%
}(\left\vert \Psi_{123}\right\rangle \left\langle \Psi_{123}\right\vert )$ and
state coefficients is given by $\left(  \rho_{12}\right)  _{i_{1}i_{2}%
j_{1}j_{2}}=\sum_{i_{3}}a_{i_{1}i_{2}i_{3}}a_{j_{1}j_{2}i_{3}}^{\ast}$.
Similarly for $\rho_{13}=Tr_{A_{2}}(\left\vert \Psi_{123}\right\rangle
\left\langle \Psi_{123}\right\vert )$, we have $\left(  \rho_{13}\right)
_{i_{1}i_{3}j_{1}j_{3}}=\sum_{i_{2}}a_{i_{1}i_{2}i_{3}}a_{j_{1}i_{2}j_{3}%
}^{\ast}$. One can verify that $C\left(  \rho_{1j}\right)  \geq0$ for ($j=2$
and $3$), while
\begin{equation}
n_{8}\left(  \rho_{1j}\right)  =\frac{1}{16}\tau_{1|2|3}^{2}\left(  \left\vert
\Psi_{123}\right\rangle \right)  ,\quad n_{12}\left(  \rho_{1j}\right)
=n_{16}\left(  \rho_{1j}\right)  =0,
\end{equation}
and
\begin{equation}
\tau_{1|23}\left(  \left\vert \Psi_{123}\right\rangle \right)  =n_{4}\left(
\rho_{12}\right)  +n_{4}\left(  \rho_{13}\right)  . \label{tan3n4}%
\end{equation}
From Eq. (\ref{2tanrho}), the two-tangle of the state $\rho_{1j}$ reads as%
\begin{equation}
\tau_{1|j}^{2}\left(  \rho_{1j}\right)  =n_{4}\left(  \rho_{1j}\right)
-\frac{1}{2}\tau_{1|2|3}\left(  \left\vert \Psi_{123}\right\rangle \right)
;\quad\left(  j=2,3\right)  . \label{2tan3qubit}%
\end{equation}
Substituting the value of coefficients $n_{4}\left(  \rho_{1j}\right)  $ from
Eq (\ref{2tan3qubit}) into Eq. (\ref{tan3n4}), the tangles for $\left\vert
\Psi_{123}\right\rangle $ satisfy the constraint (CKW inequality):%
\begin{equation}
\tau_{1|23}\left(  \left\vert \Psi_{123}\right\rangle \right)  =\tau_{1|2}%
^{2}\left(  \rho_{12}\right)  +\tau_{1|3}^{2}\left(  \rho_{13}\right)
+\tau_{1|2|3}\left(  \left\vert \Psi_{123}\right\rangle \right)  \text{.}
\label{ckw}%
\end{equation}
In other words, with qubit $A_{1}$ as focus qubit the sum of two-tangles and
three-way correlations in $\left\vert \Psi_{123}\right\rangle $ is equal to
$\tau_{1|23}\left(  \left\vert \Psi_{123}\right\rangle \right)  $. Analogous
relations can be found by taking $A_{2}$ or $A_{3}$ as the focus qubit. It
implies that stronger the entanglement of a qubit pair in a three-qubit pure
state, the weaker is entanglement of the pair with the rest of the system.
This also implies that if three-way correlations are maximal that is
$\tau_{1|2|3}\left(  \left\vert \Psi_{123}\right\rangle \right)
=1=\tau_{1|23}\left(  \left\vert \Psi_{123}\right\rangle \right)  $, then
$\tau_{1|2}^{2}\left(  \rho_{12}\right)  =\tau_{1|3}^{2}\left(  \rho
_{13}\right)  =0$.

\section{One-tangle of a Four-qubit pure State\label{fourstate}}

In this section, we consider the case where two-qubit state is a marginal
state of four-qubit composite system in a pure state. An understanding of
distribution of quantum correlations in a pure state with more than\ three
qubits is a fascinating challenge. Our main objective is to find the relation
between one-tangle of the state with qubit $A_{1}$ as the focus qubit,
coefficients $n_{4}\left(  \rho_{1j}\right)  $ and $n_{8}\left(  \rho
_{1j}\right)  $ ($j=2$ to $4$). To facilitate the calculation, the formalism
of determinants of negativity fonts is used to express $n_{4}\left(  \rho
_{1j}\right)  $ and $n_{8}\left(  \rho_{1j}\right)  $ in terms of two-qubit,
three-qubit and four-qubit unitary invariant combinations of state
coefficients. For more on definition and physical meaning of determinants of
negativity fonts, please refer to section (VI) of ref. \cite{shar16}. A
general four-qubit pure state reads as%
\begin{equation}
\left\vert \Psi_{1234}\right\rangle =\sum_{i_{1},i_{2},i_{3},i_{4}}%
a_{i_{1}i_{2}i_{3}i_{4}}\left\vert i_{1}i_{2}i_{3}i_{4}\right\rangle
,\quad\left(  i_{m}=0,1\right)  , \label{4state}%
\end{equation}
where the state coefficients $a_{i_{1}i_{2}i_{3}i_{4}}$ are complex numbers.
The indices $i_{1}$, $i_{2}$, $i_{3}$, $i_{4}$ refer, respectively, to the
state of qubits $A_{1}$, $A_{2}$, $A_{3}$, and $A_{4}$. Taking qubit $A_{1}$
as the focus qubit, for the purpose of this article the determinants of
negativity fonts of $\left\vert \Psi_{1234}\right\rangle $ are defined as
$D_{\left(  A_{3}\right)  _{i_{3}}\left(  A_{4}\right)  _{i_{4}}}%
^{00}=a_{00i_{3}i_{4}}a_{11i_{3}i_{4}}-a_{10i_{3}i_{4}}a_{01i_{3}i_{4}}$
(two-way), $D_{\left(  A_{2}\right)  _{i_{2}}\left(  A_{4}\right)  _{i_{4}}%
}^{00}=a_{0i_{2}0i_{4}}a_{1i_{2}1i_{4}}-a_{1i_{2}0i_{4}}a_{0i_{2}1i_{4}}$
(two-way), $D_{\left(  A_{2}\right)  _{i_{2}}\left(  A_{3}\right)  _{i_{3}}%
}^{00}=a_{0i_{2}i_{3}0}a_{1i_{2}i_{3}1}-a_{1i_{2}i_{3}0}a_{0i_{2}i_{3}1}$
(two-way), $D_{\left(  A_{4}\right)  _{i_{4}}}^{00i_{3}}=a_{00i_{3}i_{4}%
}a_{11,i_{3}\oplus1,i_{4}}-a_{10i_{3}i_{4}}a_{01,i_{3}\oplus1,i_{4}}$
(three-way), $D_{\left(  A_{3}\right)  _{i_{3}}}^{00i_{4}}=a_{00i_{3}i_{4}%
}a_{11i_{3},i_{4}\oplus1}-a_{10i_{3}i_{4}}a_{01i_{3},i_{4}\oplus1}$
(three-way), $D_{\left(  A_{2}\right)  _{i_{2}}}^{00i_{4}}=a_{0i_{2}0i_{4}%
}a_{1i_{2}1i_{4}\oplus1}-a_{1i_{2}0i_{4}}a_{0i_{2}1i_{4}\oplus1}$ (three-way),
and $D^{00i_{3}i_{4}}=a_{00i_{3}i_{4}}a_{11,i_{3}\oplus1,i_{4}\oplus
1}-a_{10i_{3}i_{4}}a_{01,i_{3}\oplus1,i_{4}\oplus1}$ (four-way). All these
determinants correspond to the set of $D_{1jIJ}$ obtained by substituting
$j=\left(  2,3,4\right)  $ and $I\equiv\left\{  00,10,01,11\right\}  $, in Eq.
(\ref{d1jIJ}). To understand what does the determinant of a four-way
negativity font represent, consider the state%
\begin{equation}
\left\vert \Psi\right\rangle =a_{0000}\left\vert 0000\right\rangle
+a_{1000}\left\vert 1000\right\rangle +a_{0111}\left\vert 0111\right\rangle
+a_{1111}\left\vert 1111\right\rangle ,
\end{equation}
with $D^{0000}=a_{0000}a_{1111}-a_{1000}a_{0111}$. It is easily verified that
taking negativity of partial transpose of $\left\vert \Psi\right\rangle $ with
respect to qubit $A_{1}$ as the entanglement measure, the entanglement of
qubit $A_{1}$ with the three remaining qubits due to four-way correlations is
$4\left\vert D^{0000}\right\vert $.

Matrix elements of the state $\rho_{12}=Tr_{A_{3}A_{4}}(\left\vert \Psi
_{1234}\right\rangle \left\langle \Psi_{1234}\right\vert )$ are given by%
\begin{equation}
\left(  \rho_{12}\right)  _{i_{1}i_{2}j_{1}j_{2}}=\sum_{i_{3}i_{4}}%
a_{i_{1}i_{2}i_{3}i_{4}}a_{j_{1}j_{2}i_{3}i_{4}}^{\ast}. \label{mero12}%
\end{equation}
We use Eq. (\ref{mero12}) to express the characteristic polynomial of
$\rho_{12}\widetilde{\rho_{12}}$ in terms of state coefficients of $\left\vert
\Psi_{1234}\right\rangle $ and identify the coefficient $n_{4}\left(
\rho_{12}\right)  $. Similarly one can obtain the coefficients $n_{4}\left(
\rho_{13}\right)  $ and $n_{4}\left(  \rho_{14}\right)  $ from the states
$\rho_{13}=Tr_{A_{2}A_{4}}(\left\vert \Psi_{1234}\right\rangle \left\langle
\Psi_{1234}\right\vert )$ and $\rho_{14}=Tr_{A_{2}A_{3}}(\left\vert
\Psi_{1234}\right\rangle \left\langle \Psi_{1234}\right\vert )$, respectively.
Expression for $n_{4}\left(  \rho_{1j}\right)  $ ($j=2$ to $4$) in terms of
determinants of negativity fonts is given by Eq. (\ref{n4form}) in subsection
\ref{2} of Appendix \ref{B}.

One-tangle, $\tau_{1|234}\left(  \left\vert \Psi_{1234}\right\rangle \right)
=4\det\left(  \rho_{1}\right)  $ with $\rho_{1}=Tr_{A_{2}A_{3}A_{4}%
}(\left\vert \Psi_{1234}\right\rangle \left\langle \Psi_{1234}\right\vert )$,
quantifies the entanglement of qubit $A_{1}$ with qubits $A_{2}A_{3}A_{4}$.
Using Eq. (\ref{1tann4N}), it is easily verified that one-tangle satisfies the
relation,%
\begin{equation}
\tau_{1|234}=\sum_{j=2}^{4}n_{4}\left(  \rho_{1j}\right)  -\frac{1}{2}\left(
\tau_{1|2|3|4}^{(0)}\right)  ^{2}.\label{1tann4}%
\end{equation}
In Eq. (\ref{1tann4}), the four-qubit tangle $\tau_{1|2|3|4}^{(0)}$ is defined
as%
\begin{equation}
\tau_{1|2|3|4}^{(0)}=2\left\vert D^{0000}+D^{0011}-D^{0010}-D^{0001}%
\right\vert .\label{0fourtan}%
\end{equation}
It is known to detect GHZ-like entanglement of a four-qubit state, vanishes on
a W-like state of four qubits, however, fails to vanish on product of
two-qubit entangled states. Four-qubit invariant of degree two
\begin{equation}
I_{4,2}=D^{0000}+D^{0011}-D^{0010}-D^{0001}%
\end{equation}
is the same as degree-two invariant H of ref. \cite{luqu03}.

\section{Three-tangles of a four-qubit state and unitary invariants of degree
eight}

To decipher the nature of correlations represented by $n_{4}\left(  \rho
_{1j}\right)  $ and $n_{8}\left(  \rho_{1j}\right)  $, we write the
characteristic polynomial of matrix $\rho_{1j}\widetilde{\rho_{1j}}$ in terms
of the state coefficients of $\left\vert \Psi_{1234}\right\rangle $ and
identify the coefficients $n_{4}\left(  \rho_{1j}\right)  $, $n_{8}\left(
\rho_{1j}\right)  $, $n_{12}\left(  \rho_{1j}\right)  $ and $n_{16}\left(
\rho_{1j}\right)  $. A rather lengthy analytical calculation reveals that when
two-qubit state $\rho_{1j}$ is a marginal state of $\left\vert \Psi
_{1234}\right\rangle $ then the coefficient $n_{4}\left(  \rho_{1j}\right)  $
is a sum of squares of moduli of two-qubit invariants while the coefficient
$n_{8}\left(  \rho_{1j}\right)  $ is a sum of three-qubit invariants.
Expressions for $n_{4}\left(  \rho_{1j}\right)  $ (Eq. (\ref{n4form})) and
$n_{8}\left(  \rho_{1j}\right)  $ (Eq. (\ref{n8form})) are given in subsection
\ref{2} of Appendix \ref{B}. The coefficient $n_{8}\left(  \rho_{1j}\right)  $
can, in turn, be rewritten as a sum of four-qubit unitary invariant
combinations of three-qubit invariants. This section deals with the relation
between the three-tangle of a given triple in a four-qubit pure state and the
corresponding four-qubit invariant. It is shown in the following section, that
the coefficient $n_{8}\left(  \rho_{1j}\right)  ,$ $\left\{  j=2,3,4\right\}
$, is a function of two of the three-tangles $\tau_{1|j|k}\left(  \rho
_{1jk}\right)  ,$ $\left\{  k=2,3,4:k\neq j\right\}  $ and four-tangles.

It has been shown in our earlier works that given a three-qubit marginal state
of a four-qubit state, the upper bound on three-tangle \cite{shar17} depends
on a specific unitary invariant \cite{shar16} of the pure four-qubit state and
genuine four-tangle of the state \cite{shar14,shar16}. Using the definition of
three-tangle of a mixed state (Eq. (\ref{mix3tan})), we identify the
four-qubit invariant which contains the three-tangle of a mixed state. For the
state $\left\vert \Psi_{1234}\right\rangle $, the three-qubit invariants
corresponding to $I_{3,4}\left(  \left\vert \Psi_{123}\right\rangle \right)  $
(Eq. (\ref{3invariant})) read as%
\begin{equation}
I_{A_{4}}^{4,0}=\left(  D_{\left(  A_{4}\right)  _{0}}^{000}+D_{\left(
A_{4}\right)  _{0}}^{001}\right)  ^{2}-4D_{\left(  A_{3}\right)  _{0}\left(
A_{4}\right)  _{0}}^{00}D_{\left(  A_{3}\right)  _{1}\left(  A_{4}\right)
_{0}}^{00}.
\end{equation}
and
\begin{equation}
I_{A_{4}}^{0,4}=\left(  D_{\left(  A_{4}\right)  _{1}}^{000}+D_{\left(
A_{4}\right)  _{1}}^{001}\right)  ^{2}-4D_{\left(  A_{3}\right)  _{0}\left(
A_{4}\right)  _{1}}^{00}D_{\left(  A_{3}\right)  _{1}\left(  A_{4}\right)
_{1}}^{00}.
\end{equation}
Here superscript in $I_{A_{4}}^{4,0}$ indicates that it is a three-qubit
invariant of degree ($4+0$) that is each term is a product of four of the
state coefficients, all of which have $i_{4}=0$, and none of them contains a
state coefficient with $i_{4}=1$. Likewise, $I_{A_{4}}^{0,4}$ is a three-qubit
invariant with each term being a product of four state coefficients all of
which have $i_{4}=1$. The superscript contains information about the
transformation properties of the invariant under the action of a unitary
U$^{4}=\frac{1}{\sqrt{1+\left\vert x\right\vert ^{2}}}\left[
\begin{array}
[c]{cc}%
1 & -x^{\ast}\\
x & 1
\end{array}
\right]  $ on qubit $A_{4}$. One can verify that $I_{A_{4}}^{4,0}\left(
U^{4}\left\vert \Psi_{1234}\right\rangle \right)  $ is a function of
three-qubit invariants contained in the set $\left\{  I_{A_{4}}^{4-m,m}%
:m=0,4\right\}  $. Here superscript in element $I_{A_{4}}^{4-m,m}$ indicates
that it is a three-qubit invariant of degree four such that each term is a
product of $\left(  4-m\right)  $ state coefficients with $i_{4}=0$, and $m$
state coefficient with $i_{4}=1$.

The form of elements of the set in terms of determinants of negativity fonts
is given in subsection \ref{3} of Appendix \ref{B}. Four-qubit invariant that
quantifies the three-way and genuine four-way correlations \cite{shar16} of
triple $A_{1}A_{2}A_{3}$, reads as%
\begin{equation}
N_{4,8}^{(123)}=\left\vert I_{A_{4}}^{4,0}\right\vert ^{2}+4\left\vert
I_{A_{4}}^{3,1}\right\vert ^{2}+6\left\vert I_{A_{4}}^{2,2}\right\vert
^{2}+4\left\vert I_{A_{4}}^{1,3}\right\vert ^{2}+\left\vert I_{A_{4}}%
^{0,4}\right\vert ^{2},
\end{equation}
whereas the degree-eight invariant that measures genuine four-way entanglement
of the state $\left\vert \Psi_{1234}\right\rangle $ is given by%
\begin{equation}
I_{4,8}=3\left(  I_{A_{4}}^{2,2}\right)  ^{2}-4I_{A_{4}}^{3,1}I_{A_{4}}%
^{1,3}+I_{A_{4}}^{4,0}I_{A_{4}}^{0,4}.
\end{equation}
First subscript in $N_{4,8}^{(123)}$ or $I_{4,8}$ indicates that it is a
four-qubit invariant while the second subscript indicates the degree of the invariant.

On the other hand, for the mixed state $\rho_{123}=Tr_{A_{4}}(\left\vert
\Psi_{1234}\right\rangle \left\langle \Psi_{1234}\right\vert )=\sum
\limits_{i=0,1}p_{i}\left\vert \phi_{123}^{\left(  i\right)  }\right\rangle
\left\langle \phi_{123}^{\left(  i\right)  }\right\vert $, the three tangle
(Eq. (\ref{mix3tan})) is given by%
\begin{equation}
\left[  \tau_{1|2|3}\left(  \rho_{123}\right)  \right]  ^{\frac{1}{2}}%
=2\min_{\left\{  p_{i},\left\vert \phi_{123}^{\left(  i\right)  }\right\rangle
\right\}  }\left\{  \left\vert I_{A_{4}}^{4,0}\right\vert ^{\frac{1}{2}%
}+\left\vert I_{A_{4}}^{0,4}\right\vert ^{\frac{1}{2}}\right\}  .
\label{tan123}%
\end{equation}
It is known from ref. \cite{shar17} that the upper bound on $\tau
_{1|2|3}\left(  \rho_{123}\right)  $, is given by%
\begin{equation}
\tau_{1|2|3}^{up}\left(  \rho_{123}\right)  =\sqrt{16N_{4,8}^{(123)}-\frac
{1}{6}\left(  \tau_{1|2|3|4}^{(1)}\right)  ^{2}},
\end{equation}
where $\tau_{1|2|3|4}^{(1)}=\sqrt{16\left\vert 12I_{4,8}\right\vert }$ is the
genuine four-tangle defined in refs. \cite{shar14,shar16}.

In general, for a selection of three qubits $A_{1}A_{j}A_{k}$, where $j=2$ to
$4$ and $k=2$ to $4$, with the appropriate set of three-qubit invariants
$\left\{  I_{A_{i}}^{4-m,m}\left(  \left\vert \Phi_{i}\right\rangle \right)
:m=0,4\text{, }i\neq j\neq k\right\}  $, degree-eight invariant $N_{4,8}%
^{(1jk)}$ and three-tangle $\tau_{1|j|k}\left(  \rho_{1jk}\right)  $ satisfy
the inequality%
\begin{equation}
\sqrt{16N_{4,8}^{(1jk)}-\frac{1}{6}\left(  \tau_{1|2|3|4}^{(1)}\right)  ^{2}%
}\geq\tau_{1|j|k}\left(  \rho_{1jk}\right)  . \label{3tanrho1jk}%
\end{equation}
In case $\tau_{1|j|k}\left(  \rho_{1jk}\right)  =0$, $16N_{4,8}^{(1jk)}%
=\frac{1}{6}\left(  \tau_{1|2|3|4}^{(1)}\right)  ^{2}$. Expressions for
$N_{4,8}^{(1jk)}$ and $\tau_{1|2|3|4}^{(1)}$ are given in subsection \ref{4}
of Appendix \ref{B}.

\section{What does coefficient $n_{8}\left(  \rho_{1j}\right)  $ represent?}

An analytical calculation reveals that for the two-qubit state, $\rho
_{12}=Tr_{A_{3}A_{4}}(\left\vert \Psi_{1234}\right\rangle \left\langle
\Psi_{1234}\right\vert )$, the coefficient $n_{8}\left(  \rho_{12}\right)  $
is a sum of four-qubit invariants. Two of these four-qubit invariants are
$N_{4,8}^{\left(  123\right)  }$ and $N_{4,8}^{\left(  124\right)  }$. The
coefficient $n_{8}\left(  \rho_{12}\right)  $ also contains contribution from
$\left\vert 3\left(  I_{4,2}\right)  ^{2}-P_{12}\right\vert $, where
$P_{1j},(j=2$ to $4)\ $are already known from earlier works on polynomial
invariants \cite{shar10}. Subscripts on $P_{1j}$ refer to the pair of qubits
$A_{1}A_{j}$. The invariant $P_{1j}$ is non zero if the qubit pair $A_{1}%
A_{j}$ is entangled to the rest of the system in the four-qubit pure state
$\left\vert \Psi_{1234}\right\rangle $. Detailed form of these invariants in
terms of determinants of negativity fonts is given in subsection \ref{5} of
appendix \ref{B}. It is easily verified that $P_{1j},(j=2$ to $4)$ are not
independent invariants because%
\begin{equation}
P_{12}+P_{13}+P_{14}=3\left(  I_{4,2}\right)  ^{2}.
\end{equation}
The exact expression for coefficient $n_{8}\left(  \rho_{12}\right)  $ reads
as%
\begin{align}
n_{8}\left(  \rho_{12}\right)   &  =N_{4,8}^{\left(  123\right)  }%
+N_{4,8}^{\left(  124\right)  }\nonumber\\
&  +\frac{1}{24}\left\vert 3\left(  I_{4,2}\right)  ^{2}-P_{12}\right\vert
^{2}+M_{4,8}\left(  \rho_{12}\right)  , \label{n8rho12}%
\end{align}
where $M_{4,8}\left(  \rho_{12}\right)  $ is a sum of three-qubit invariants.
Expression for $M_{4,8}\left(  \rho_{1j}\right)  $ is also given in subsection
\ref{5} of appendix \ref{B}. Form of each term in $M_{4,8}\left(  \rho
_{12}\right)  $ reveals that this four-qubit invariant is non-zero only on a
four-qubit state. Similarly the coefficients $n_{8}\left(  \rho_{13}\right)  $
and $n_{8}\left(  \rho_{14}\right)  $ read as%
\begin{align}
n_{8}\left(  \rho_{13}\right)   &  =N_{4,8}^{\left(  123\right)  }%
+N_{4,8}^{\left(  134\right)  }\nonumber\\
&  +\frac{1}{24}\left\vert 3\left(  I_{4,2}\right)  ^{2}-P_{13}\right\vert
^{2}+M_{4,8}\left(  \rho_{13}\right)  , \label{n8rho13}%
\end{align}
and%
\begin{align}
n_{8}\left(  \rho_{14}\right)   &  =N_{4,8}^{\left(  124\right)  }%
+N_{4,8}^{\left(  134\right)  }\nonumber\\
&  +\frac{1}{24}\left\vert 3\left(  I_{4,2}\right)  ^{2}-P_{14}\right\vert
^{2}+M_{4,8}\left(  \rho_{14}\right)  , \label{n8rho14}%
\end{align}

A comparison of $n_{8}\left(  \rho_{1j}\right)  $ $\left\{  j=2,3,4\right\}  $
with the upper bound on three-tangles from Eq. (\ref{3tanrho1jk}) shows that
$n_{8}\left(  \rho_{1j}\right)  $ is a function of two of the three-tangles
$\tau_{1|j|k}\left(  \rho_{1jk}\right)  $ such that%
\begin{equation}
4n_{8}\left(  \rho_{1j}\right)  =\frac{1}{4}\sum_{k=2;k\neq j}^{4}\tau
_{1|j|k}^{2}\left(  \rho_{1jk}\right)  +\delta_{1j},\label{n81jc}%
\end{equation}
where%
\begin{equation}
\delta_{1j}\geq\frac{1}{12}\left(  \tau_{1|2|3|4}^{(1)}\right)  ^{2}+\frac
{1}{8}\left(  \tau_{1|2|3|4}^{(2)}\left(  \rho_{1j}\right)  \right)
^{2}+\frac{3}{32}\left(  \tau_{1|2|3|4}^{(3)}\left(  \rho_{1j}\right)
\right)  ^{2}.\label{delta1j}%
\end{equation}
Here we have defined four-tangles, $\tau_{1|2|3|4}^{(2)}\left(  \rho
_{1j}\right)  =\sqrt{32M_{4,8}\left(  \rho_{1j}\right)  }$ and $\tau
_{1|2|3|4}^{(3)}\left(  \rho_{1j}\right)  =\left\vert 4\left(  I_{4,2}\right)
^{2}-\frac{4}{3}P_{1j}\right\vert $. The quantity $\delta_{1j}$ is a function
of four-way correlations. Eq. (\ref{n81jc}) represents an interesting
condition on how three-way and four-way correlations are shared by qubits. For
example, if $4n_{8}\left(  \rho_{12}\right)  =\frac{1}{4}$ and $\tau
_{1|2|3}^{2}\left(  \rho_{123}\right)  =1$, then $\tau_{1|2|4}^{2}\left(
\rho_{124}\right)  =0$ and $\delta_{12}=0$.

\section{Constraint on Three-tangles and Four-tangles}

We may note that the sum of degree eight coefficients constrains the amount of
three-way and four-way correlations in a four-qubit state. If $N_{4,8}%
^{\left(  123\right)  }$, $N_{4,8}^{\left(  124\right)  }$ as well as
$N_{4,8}^{\left(  134\right)  }$ are non-zero, then the sum $4\sum_{j=2}%
^{4}n_{8}\left(  \rho_{1j}\right)  $ is found to satisfy the constraint%
\begin{equation}
4\sum_{j=2}^{4}n_{8}\left(  \rho_{1j}\right)  -\frac{1}{2}\left(  \tau
_{1|2|3}^{2}\left(  \rho_{123}\right)  +\tau_{1|2|4}^{2}\left(  \rho
_{124}\right)  +\tau_{1|3|4}^{2}\left(  \rho_{134}\right)  \right)
=\sum_{j=2}^{4}\delta_{1j} \label{sum4n8c}%
\end{equation}
where the residue $\sum_{j=2}^{4}\delta_{1j}$ is a function of four-way
correlations characterizing the pure state $\left\vert \Psi_{1234}%
\right\rangle $ and reads as%
\begin{equation}
\sum_{j=2}^{4}\delta_{1j}\geq\frac{1}{4}\tau_{1|2|3|4}^{(1)}+\frac{3}{32}%
\sum_{j=2}^{4}\left(  \tau_{1|2|3|4}^{(3)}\left(  \rho_{1j}\right)  \right)
^{2}+\frac{1}{8}\sum_{j=2}^{4}\left(  \tau_{1|2|3|4}^{(2)}\left(  \rho
_{1j}\right)  \right)  ^{2}. \label{sumdelta1j}%
\end{equation}
By construction $\tau_{1|2|3|4}^{(3)}\left(  \rho_{1j}\right)  $ is non-zero
if and only if the qubit pair $A_{1}A_{j}$ in the four-qubit state is
entangled to the rest of the system. It is easily verified that $\left\vert
P^{A_{1}A_{2}}\right\vert =\left\vert P^{A_{3}A_{4}}\right\vert $, $\left\vert
P^{A_{1}A_{3}}\right\vert =\left\vert P^{A_{2}A_{4}}\right\vert $, and
$\left\vert P^{A_{1}A_{4}}\right\vert =\left\vert P^{A_{2}A_{3}}\right\vert $,
as such, $\sum_{j=2}^{4}\left(  \tau_{1|2|3|4}^{(3)}\left(  \rho_{1j}\right)
\right)  ^{2}$ does not depend on the choice of focus qubit.

Four tangles $\tau_{1|2|3|4}^{(0)}$ , $\tau_{1|2|3|4}^{(1)}$, $\tau
_{1|2|3|4}^{(2)}\left(  \rho_{1j}\right)  $ and $\tau_{1|2|3|4}^{(3)}\left(
\rho_{1j}\right)  $ are invariant with respect to a local unitary on anyone of
the four qubits. Just as a three-tangle is defined only on states with
$N\geq3$, four-tangles are defined only on states with $N\geq4$. Here,
$\tau_{1|2|3|4}^{(0)}$ is defined in terms of a degree-two invariant, while
$\tau_{1|2|3|4}^{(1)}$ is a function of a single four-qubit invariant of
degree eight. The genuine four-tangle $\tau_{1|2|3|4}^{(1)}>0$ implies that
each one of the qubits is entangled to the three remaining qubits due to
four-way correlations. A measurement on one of the four-qubits of a pure
four-qubit state completely destroys the entanglement quantified by
$\tau_{1|2|3|4}^{(1)}$. Four-tangle $\tau_{1|2|3|4}^{(1)}$ is the analog of
three-tangle for three-qubit states. On a W-state of four qubits we have
$\tau_{1|2|3|4}^{(1)}=0$.

To understand the role of four-tangle, $\tau_{1|2|3|4}^{(2)}\left(
\rho\right)  $, we consider a simple four-qubit state on which $\tau
_{1|2|3|4}^{(2)}\left(  \rho_{12}\right)  \neq0$ that is%
\begin{equation}
\left\vert \chi\right\rangle =a_{0000}\left\vert 0000\right\rangle
+a_{1101}\left\vert 1101\right\rangle +a_{1110}\left\vert 1110\right\rangle .
\label{psiexample}%
\end{equation}
One can verify that on $\left\vert \chi\right\rangle $, three tangles take
value $\tau_{1|2|3}\left(  \rho_{123}\right)  =4\left\vert \left(
a_{0000}a_{1110}\right)  ^{2}\right\vert $, $\tau_{1|2|4}\left(  \rho
_{124}\right)  =4\left\vert \left(  a_{0000}a_{1101}\right)  ^{2}\right\vert $
and%
\begin{equation}
\left(  \tau_{1|2|3|4}^{(2)}\left(  \rho_{12}\right)  \right)  ^{2}%
=4\tau_{1|2|3}\tau_{1|2|4}.
\end{equation}
On the other hand $\tau_{1|2|3|4}^{(2)}\left(  \rho_{13}\right)
=\tau_{1|2|3|4}^{(2)}\left(  \rho_{14}\right)  =0$. Four-tangle $\tau
_{1|2|3|4}^{(2)}\left(  \rho_{1j}\right)  $ is a sum of nine three-qubit
invariants of degree eight. It contains contributions from products of
three-tangles of underlying three-qubit subsystems.

\section{Constraints on Tangles of a four-qubit state}

In the case of a three-qubit pure state monogamy relation is a relation
between degree-four functions of state coefficients that is one-tangle
$\tau_{1|23}\left(  \left\vert \Psi_{123}\right\rangle \right)  $, square of
two-tangle $\tau_{1|j}^{2}\left(  \rho_{1j}\right)  $ and three-tangle
$\tau_{1|2|3}\left(  \left\vert \Psi_{123}\right\rangle \right)  $. Genuine
four-way entanglement \cite{shar14,shar16}, however, is quantified by a
degree-eight function of state coefficients. Consequently, we have distinct
sets of constraints to be satisfied by degree-four and degree-eight
entanglement measures of correlations of a four-qubit state. A constraint on
one-tangle and two-tangles is obtained by subtracting the sum of two tangles
from Eq. (\ref{1tann4}) that is%
\begin{align}
S_{1}  &  =\sum_{j=2}^{4}\left(  n_{4}\left(  \rho_{1j}\right)  -\tau
_{1|j}^{2}\left(  \rho_{1j}\right)  \right)  -\frac{1}{2}\left(
\tau_{1|2|3|4}^{(0)}\right)  ^{2}\nonumber\\
&  =\tau_{1|234}-\sum_{j=2}^{4}\tau_{1|j}^{2}\left(  \rho_{1j}\right)  ,
\label{S1}%
\end{align}
where\ $S_{1}$ represents three- and four-way correlations.

The state $\rho_{1j}$ $\left\{  j=2,3,4\right\}  $ being a reduced state of
$\rho_{1jk}$ $\left\{  k=2,3,4:k\neq j\right\}  $ contains information about
two-tangle $\tau_{1|j}\left(  \rho_{1j}\right)  $, two of the three tangles
$\tau_{1|j|k}\left(  \rho_{1jk}\right)  $, as well as four-way correlations.
If for a two-qubit marginal state $\rho_{1j}\left\{  j=2,3,4\right\}  $ of
$\left\vert \Psi_{1234}\right\rangle $, then the relation analogous to Eq.
(\ref{2tanrho}) reads as%
\begin{equation}
n_{4}\left(  \rho_{1j}\right)  =\tau_{1|j}^{2}\left(  \rho_{1j}\right)
+\sqrt{4n_{8}\left(  \rho_{1j}\right)  +\chi^{\pm}\left(  \rho_{1j}\right)  },
\label{n41j}%
\end{equation}
with $\chi^{\pm}\left(  \rho_{1j}\right)  $ defined as in (Eqs.
(\ref{chiroplus}) and (\ref{chirominus})) that is%
\begin{equation}
\chi^{+}\left(  \rho_{1j}\right)  =8\sqrt{n_{16}\left(  \rho_{1j}\right)
}+8\sqrt{f_{16}\left(  \rho_{1j}\right)  }\text{ ,} \label{chi1jplus}%
\end{equation}
and%
\begin{align}
\chi^{-}\left(  \rho_{1j}\right)   &  =8\sqrt{n_{16}\left(  \rho_{1j}\right)
}-8\sqrt{f_{16}\left(  \rho_{1j}\right)  }\nonumber\\
&  +2n_{4}\left(  \rho_{1j}\right)  \left\vert C\left(  \rho_{1j}\right)
\right\vert ^{2}-\left\vert C\left(  \rho_{1j}\right)  \right\vert ^{4}\text{
}. \label{chi1jminus}%
\end{align}
Recalling that for a two-qubit state $n_{4}\left(  \rho_{1j}\right)
=tr\left(  \rho_{1j}\widetilde{\rho}_{1j}\right)  $ is a calculable quantity,
we obtain a set of three conditions to be satisfied by measures of two-way,
three-way and four-way correlations. Substituting for coefficients
$n_{8}\left(  \rho_{1j}\right)  $ from Eqs. (\ref{n81jc}) into Eq.
(\ref{n41j}), we obtain the constraints:%
\begin{equation}
\left(  n_{4}\left(  \rho_{1j}\right)  -\tau_{1|j}^{2}\left(  \rho
_{1j}\right)  \right)  ^{2}-\frac{1}{4}\sum_{k=2,k\neq j}^{4}\tau_{1|j|k}%
^{2}\left(  \rho_{1jk}\right)  =\Delta_{1j}, \label{n41jc}%
\end{equation}
where $\Delta_{1j}=\delta_{1j}+\chi^{\pm}\left(  \rho_{1j}\right)  $ and
$j=2,3,4$. Here $\delta_{1j}\geq0$ (Eq. (\ref{delta1j})), $n_{16}\left(
\rho_{1j}\right)  $ $\geq0$ and $f_{16}\left(  \rho_{1j}\right)  \geq0$
\ref{f16}) is valid. If $C\left(  \rho_{1j}\right)  \leq0$, then $\tau
_{1|j}\left(  \rho_{1j}\right)  =0$ and Eq. (\ref{n41jc}) reduces to%
\begin{equation}
n_{4}^{2}\left(  \rho_{1j}\right)  -\frac{1}{4}\sum_{k=2,k\neq j}^{4}%
\tau_{1|j|k}^{2}\left(  \rho_{1jk}\right)  =\Delta_{1j}. \label{n41jcnegative}%
\end{equation}
Using Eq. (\ref{n4form}) and the definition of $\tau_{1|j|k}\left(  \rho
_{1jk}\right)  $ (Eq. (\ref{mix3tan})), one may verify that for $\tau
_{1|j}\left(  \rho_{1j}\right)  =0$, $n_{4}^{2}\left(  \rho_{1j}\right)
\geq\frac{1}{4}\sum_{k=2,k\neq j}^{4}\tau_{1|j|k}^{2}\left(  \rho
_{1jk}\right)  $. As such, $\Delta_{1j}\geq0$ is satisfied independent of the
value of two-tangle. The quantity $\Delta_{1j}$ represents four-way
correlations involving the qubit pair $A_{1}A_{j}$ and the two remaining
qubits of the four-qubit state. However, $\chi^{-}\left(  \rho_{1j}\right)  $
may take negative values.

We notice that%
\begin{equation}
\sum_{j=2}^{4}\left(  n_{4}\left(  \rho_{1j}\right)  -\tau_{1|j}^{2}\left(
\rho_{1j}\right)  \right)  ^{2}-\frac{1}{2}\left(  \tau_{1|2|3}^{2}\left(
\rho_{123}\right)  +\tau_{1|2|4}^{2}\left(  \rho_{124}\right)  +\tau
_{1|3|4}^{2}\left(  \rho_{134}\right)  \right)  =\sum_{j=2}^{4}\Delta_{1j}
\label{sumn4c}%
\end{equation}
where $\sum_{j=2}^{4}\Delta_{1j}$ may be taken as a degree-eight measure of
residual correlations in the state $\left\vert \Psi_{1234}\right\rangle $.
Substituting for $n_{4}\left(  \rho_{1j}\right)  -\tau_{1|j}^{2}\left(
\rho_{1j}\right)  $ from Eq. (\ref{n41j}) into Eq. (\ref{S1}), the constraint
on one tangle may be written as%
\begin{equation}
\tau_{1|234}+\frac{1}{2}\left(  \tau_{1|2|3|4}^{(0)}\right)  ^{2}-\sum
_{j=2}^{4}\tau_{1|j}^{2}\left(  \rho_{1j}\right)  =\sum_{j=2}^{4}\sqrt
{\frac{1}{4}\sum_{k=2,k\neq j}^{4}\tau_{1|j|k}^{2}\left(  \rho_{1jk}\right)
+\Delta_{1j}}, \label{1tanc1}%
\end{equation}
where the right hand side is a function of three-tangles of marginal
three-qubit states as well as four-tangles of the pure four-qubit state. It is
important to note that Eq. (\ref{1tanc1}) is a relation between degree-four
terms on left hand side and square root of sum of degree-eight terms on right
hand side. Four-qubit states also satisfy the constraint on one-tangle
reported in Eq. (47) of ref. (\cite{shar18}) which involves only degree-four
invariants. In that case, the contribution to one-tangle from three-qubit
correlations due to the triple $A_{1}A_{j}A_{k}$ is found to vary between
$\frac{1}{2}\tau_{1|j|k}\left(  \rho_{1jk}\right)  $ and $\tau_{1|j|k}\left(
\rho_{1jk}\right)  $, which is consistent with Eq. (\ref{1tanc1}).
Furthermore, on a state which is a product state of a three-qubit generic
state with a single qubit, Eqs. (\ref{n41jc}) and (\ref{1tanc1}) reduce to
corresponding relations for three-qubits with the values of indices $j$ and
$k$ restricted to $2$ and $3$ that is%
\begin{equation}
n_{4}\left(  \rho_{1j}\right)  -\tau_{1|j}^{2}\left(  \rho_{1j}\right)
=\frac{1}{2}\tau_{1|2|3}\left(  \rho_{123}\right)  ;\qquad\left(
j=2,3\right)  ,
\end{equation}
and%
\begin{equation}
\tau_{1|234}-\sum_{j=2}^{3}\tau_{1|j}^{2}\left(  \rho_{1j}\right)
=\tau_{1|2|3}\left(  \rho_{123}\right)  .
\end{equation}

Alternatively, after expanding the L. H. S of Eq. (\ref{1tanc1}) we may
rewrite the relation between tangles as%
\begin{equation}%
\begin{array}
[c]{c}%
\tau_{1|234}-\sum_{j=2}^{4}\tau_{1|j}^{2}\left(  \rho_{1j}\right)  -\frac
{1}{2}\sum_{j=2}^{4}\left(  \sum_{k=3,k>j}^{4}\tau_{1|j|k}^{2}\left(
\rho_{1jk}\right)  \right)  ^{\frac{1}{2}}\\
=\sum_{j=2}^{4}\sqrt{\Delta_{1j}}\left(  1-f_{1j}\right)  -\frac{1}{2}\left(
\tau_{1|2|3|4}^{(0)}\right)  ^{2},
\end{array}
\label{1tanc2}%
\end{equation}
where $f_{1j}$ is a function of $\frac{\sqrt{\Delta_{1j}\sum_{k=3,k>j}^{4}%
\tau_{1|j|k}^{2}\left(  \rho_{1jk}\right)  }}{\sqrt{\Delta_{1j}}+\sqrt
{\sum_{k=3,k>j}^{4}\tau_{1|j|k}^{2}\left(  \rho_{1jk}\right)  }}$. Therefore
the L.H.S of equation Eq. (\ref{1tanc2} represents four-way correlations. To
sum up, the tangles characterizing a four-qubit state satisfy the constraints
represented by Eqs. (\ref{n81jc}, \ref{sum4n8c}, \ref{n41jc}, \ref{sumn4c},
\ref{1tanc1}) and Eq. (\ref{1tanc2}). In the next subsections, we consider
some examples to illustrate the validity of these constraints.

\section{Four-qubit GHZ state}

Consider the maximally entangled four-qubit GHZ state%
\begin{equation}
\left\vert GHZ\right\rangle =\frac{1}{\sqrt{2}}\left(  \left\vert
0000\right\rangle +\left\vert 1111\right\rangle \right)  \text{.} \label{ghz}%
\end{equation}
Coefficients in the characteristic polynomial of the matrix $\rho
_{1j}\widetilde{\rho}_{1j}$ are%
\begin{equation}
n_{4}\left(  \rho_{1j}\right)  =\frac{1}{2},\quad4n_{8}\left(  \rho
_{1j}\right)  =\frac{1}{4},
\end{equation}%
\begin{equation}
n_{16}\left(  \rho_{1j}\right)  =n_{12}\left(  \rho_{1j}\right)
=0;\quad(j=2,3,4).
\end{equation}
As such $\chi\left(  \rho_{1j}\right)  =0$, $4n_{8}\left(  \rho_{1j}\right)
=\Delta_{1j}$ (Eq. (\ref{n81jc} ), and $\sum_{j=2}^{4}\Delta_{1j}=\frac{3}{4}%
$(Eq. (\ref{sum4n8c})). While all two-tangles and three-tangles are zero on
this state, values of four-tangles are $\tau_{1|2|3|4}^{(0)}=\tau
_{1|2|3|4}^{(1)}=\tau_{1|2|3|4}^{(2)}\left(  \rho_{1j}\right)  =1$, and
$\tau_{1|2|3|4}^{(3)}\left(  \rho_{1j}\right)  =\frac{2}{3}$. One-tangle
satisfies the relation%
\begin{equation}
\tau_{1|234}=\sum_{j=2}^{4}n_{4}\left(  \rho_{1j}\right)  -\frac{1}{2}\left(
\tau_{1|2|3|4}^{(0)}\right)  ^{2}=1. \label{1tanghz}%
\end{equation}
and since $n_{4}\left(  \rho_{1j}\right)  =\sqrt{\Delta_{1j}}$, one tangle
represents only four-way correlations and satisfies Eq. (\ref{1tanc1}).

\section{Cluster State}

Much like the $\left\vert GHZ\right\rangle $ state, all two-tangles and three
tangles\ are zero on the maximally entangled cluster state%
\begin{equation}
\left\vert \Psi_{C}\right\rangle =\frac{1}{2}\left(  \left\vert
0000\right\rangle +\left\vert 1100\right\rangle +\left\vert 0011\right\rangle
-\left\vert 1111\right\rangle \right)  ,
\end{equation}
and $\tau_{1|234}=\tau_{1|2|3|4}^{(1)}=\tau_{1|2|3|4}^{(2)}\left(  \rho
_{12}\right)  =1$ while $\tau_{1|2|3|4}^{\left(  3\right)  }\left(  \rho
_{12}\right)  =\frac{2}{3}$. But differently from $\left\vert GHZ\right\rangle
$, $\tau_{1|2|3|4}^{(0)}$ $=\tau_{1|2|3|4}^{(2)}\left(  \rho_{13}\right)
=\tau_{1|2|3|4}^{(2)}\left(  \rho_{14}\right)  =0$ and $\tau_{1|2|3|4}%
^{\left(  3\right)  }\left(  \rho_{13}\right)  =\tau_{1|2|3|4}^{\left(
3\right)  }\left(  \rho_{14}\right)  =\frac{1}{3}$. Therefore $4n_{8}\left(
\rho_{1j}\right)  =\delta_{1j}$, $n_{4}\left(  \rho_{1j}\right)  =\sqrt
{\Delta_{1j}}$, $\chi^{-}\left(  \rho_{1j}\right)  =\Delta_{1j}-\delta_{1j}$,
, and $\tau_{1|234}=\sum_{j=2}^{4}n_{4}\left(  \rho_{1j}\right)  $. Table I
lists the coefficients $n_{d}\left(  \rho_{1j}\right)  $ for $d=4,8,12,16$
along with $\delta_{1j}$, and $\chi^{-}\left(  \rho_{1j}\right)  $, for
$j=2,3,4$ for the state $\left\vert \Psi_{C}\right\rangle $. One can verify
that four-tangles satisfy the relation (refer to Eq. (\ref{delta1j}))%
\begin{equation}%
{\displaystyle\sum_{j=2}^{4}}
\delta_{1j}=\frac{1}{4}\left(  \tau_{1|2|3|4}^{(1)}\right)  ^{2}+\frac{1}%
{8}\left(  \tau_{1|2|3|4}^{(2)}\left(  \rho_{12}\right)  \right)  ^{2}%
+\frac{3}{32}%
{\displaystyle\sum_{j=2}^{4}}
\left(  \tau_{1|2|3|4}^{(3)}\left(  \rho_{1j}\right)  \right)  ^{2}%
\text{.}\label{delta1jCluster}%
\end{equation}
\begin{table}[th]
\caption{The coefficients $n_{d}\left(  \rho_{1j}\right)  $ for $d=4,8,12,16$
along with $\Delta_{1j}$, $\delta_{1j}$, $\chi_{1j}$, $P_{1j}^{2}$ and
$\tau_{1|2|3|4}^{(2)}\left(  \rho_{1j}\right)  $ for $j=2,3,4$ for the state
$\left\vert \Psi_{C}\right\rangle $.}%
\centering
\par%
\begin{tabular}
[c]{||c||c||c||c||c||c||c||}\hline\hline
State $\left\vert \Psi_{C}\right\rangle $ & $n_{4}\left(  \rho_{1j}\right)  $
& $n_{8}\left(  \rho_{1j}\right)  $ & $n_{12}\left(  \rho_{1j}\right)  $ &
$n_{16}\left(  \rho_{1j}\right)  $ & $\delta_{1j}$ & $\chi^{-}\left(
\rho_{1j}\right)  $\\\hline\hline
$\rho_{12}$ & $\frac{1}{2}$ & $\frac{1}{16}$ & $0$ & $0$ & $\frac{1}{4}$ &
$0$\\\hline\hline
$\rho_{13}$ & $\frac{1}{4}$ & $\frac{3}{128}$ & $\frac{1}{1024}$ & $\frac
{1}{65\,536}$ & $\frac{3}{32}$ & $-\frac{1}{32}$\\\hline\hline
$\rho_{14}$ & $\frac{1}{4}$ & $\frac{3}{128}$ & $\frac{1}{1024}$ & $\frac
{1}{65\,536}$ & $\frac{3}{32}$ & $-\frac{1}{32}$\\\hline\hline
Sum & $1$ & $\frac{7}{64}$ &  &  & $\frac{7}{16}$ & $-\frac{1}{16}%
$\\\hline\hline
\end{tabular}
\end{table}

It is interesting to compare the tangles of $\left\vert \Psi_{C}\right\rangle
$ with the product of two-bell states that is%
\begin{equation}
\left\vert \Psi_{P}\right\rangle =\frac{1}{2}\left(  \left\vert
00\right\rangle +\left\vert 11\right\rangle \right)  \left(  \left\vert
00\right\rangle +\left\vert 11\right\rangle \right)  .
\end{equation}
A simple calculation shows that $\tau_{1|234}=1$, $\tau_{1|2}^{2}\left(
\rho_{12}\right)  =1$,($C_{12}=1$)$,$ $\tau_{1|3}^{2}\left(  \rho_{12}\right)
=\tau_{1|4}^{2}\left(  \rho_{12}\right)  =0$, ($C_{13}=C_{14}=-\frac{1}{2}$),
$\tau_{1|2|3|4}^{(0)}\left(  \left\vert \Psi_{P}\right\rangle \right)  =1$
while Four tangles $\tau_{1|2|3|4}^{(1)}\left(  \left\vert \Psi_{P}%
\right\rangle \right)  =\tau_{1|2|3|4}^{(2)}\left(  \rho_{1j}\right)  =0$. It
turns out that $n_{8}\left(  \rho_{1j}\right)  =\frac{3}{2^{7}}$,
$n_{12}\left(  \rho_{1j}\right)  =\frac{1}{2^{10}}$, $n_{16}\left(  \rho
_{1j}\right)  =\frac{1}{2^{16}}$ and $f_{16}\left(  \rho\right)  =\frac
{1}{2^{12}}$ for $j=3$ and $4$. Consequently, $n_{4}\left(  1j\right)
=C_{1j}^{2}$ indicating that the state does not have three or four-qubit
correlations. Recalling that%
\begin{equation}
\sum_{j=2}^{4}n_{4}\left(  1j\right)  =\tau_{1|2}^{2}\left(  \rho_{12}\right)
+\sum_{j=3}^{4}C_{1j}^{2},
\end{equation}
and one tangle satisfies the relation corresponding to Eq. (\ref{1tanc1}), we
obtain%
\begin{equation}
\sum_{j=3}^{4}C_{1j}^{2}=\frac{1}{2}\left(  \tau_{1|2|3|4}^{(0)}\left(
\left\vert \Psi_{P}\right\rangle \right)  \right)  ^{2},
\end{equation}
clarifying that $\tau_{1|2|3|4}^{(0)}\left(  \left\vert \Psi_{P}\right\rangle
\right)  $ does not quantify four-way correlations.

\section{States $L_{a,ia,\left(  ia\right)  _{2}}$}

A natural extension of CKW inequality to four-qubit states may be written as
\begin{equation}
\mathcal{\tau}_{1|234}\geq\sum\limits_{j=2}^{4}\tau_{1|j}^{2}\left(  \rho
_{1j}\right)  +\sum\limits_{\substack{\left(  j,k\right)  =2\\k>j}}^{4}%
\tau_{1|j|k}\left(  \rho_{1jk}\right)  . \label{mono1}%
\end{equation}
Regula et al. \cite{regu14} have shown that a subset of four-qubit pure states
violates the inequality of Eq. (\ref{mono1}). Based on numerical evidence, the
authors of \cite{regu14} conjecture that four-qubit tangles satisfy a modified
monogamy inequality, which for four-qubits with $A_{1}$ as focus qubit, (Eq.
(9) in ref. \cite{regu14}) reads as%
\begin{align}
\mathcal{\tau}_{1|234}\left(  \left\vert \Psi_{1234}\right\rangle \right)   &
\geq\left[  \mathcal{\tau}_{1|2}\left(  \rho_{12}\right)  \right]
^{2}+\left[  \mathcal{\tau}_{1|3}\left(  \rho_{13}\right)  \right]
^{2}+\left[  \mathcal{\tau}_{1|4}\left(  \rho_{14}\right)  \right]
^{2}\nonumber\\
&  +\left[  \mathcal{\tau}_{1|2|3}\left(  \rho_{123}\right)  \right]
^{\frac{3}{2}}+\left[  \mathcal{\tau}_{1|2|4}\left(  \rho_{124}\right)
\right]  ^{\frac{3}{2}}+\left[  \mathcal{\tau}_{1|3|4}\left(  \rho
_{134}\right)  \right]  ^{\frac{3}{2}}. \label{mono2}%
\end{align}
Here three tangles are raised to the power $\frac{3}{2}$, so that the
\textquotedblleft residual four tangle\textquotedblright\ may not become
negative. Consider the product of a three qubit entangled state with the
fourth qubit in state $\left\vert 0\right\rangle $, that is
\begin{equation}
\left\vert \Psi_{s}\right\rangle =a_{0000}\left\vert 0000\right\rangle
+a_{1110}\left\vert 1110\right\rangle ,
\end{equation}
for which $\mathcal{\tau}_{1|234}\left(  \left\vert \Psi_{1234}\right\rangle
\right)  =\mathcal{\tau}_{1|2|3}\left(  \rho_{123}\right)  =4\left\vert
a_{0000}a_{1110}\right\vert ^{2}$. The inequality of Eq. (\ref{mono2}) implies
that the state $\left\vert \Psi_{s}\right\rangle $ has a \textquotedblleft
residual four tangle\textquotedblright\ given by $\mathcal{\tau}%
_{1|234}\left(  \left\vert \Psi_{1234}\right\rangle \right)  -\left[
\mathcal{\tau}_{1|2|3}\left(  \rho_{123}\right)  \right]  ^{\frac{3}{2}}$,
which is not true. On the other hand Eq. (\ref{1tanc1}) yields $\tau
_{1|234}=\mathcal{\tau}_{1|2|3}\left(  \rho_{123}\right)  $ on the state
$\left\vert \Psi_{s}\right\rangle $, as expected.

It was also pointed out by Regula et al. \cite{regu14} that states with
particularly large violations of the inequality represented by Eq.
(\ref{mono1}) can be constructed by starting with the state $L_{abc_{2}}$ of
ref. \cite{vers02}\ with $b=c$ and additionally imposing $b=c=ia$ with
parameter $a\geq0$, that is%
\begin{align}
L_{a,ia.\left(  ia\right)  _{2}}  &  =a\left(  \frac{1+i}{2}\right)  \left(
\left\vert 0000\right\rangle +\left\vert 1111\right\rangle \right)  +a\left(
\frac{1-i}{2}\right)  \left(  \left\vert 0011\right\rangle +\left\vert
1100\right\rangle \right) \nonumber\\
&  +ia\left(  \left\vert 0101\right\rangle +\left\vert 1010\right\rangle
\right)  +\left\vert 0110\right\rangle . \label{Gaiaia}%
\end{align}

\begin{figure}[th]
\centering
\includegraphics[height=3.5in]{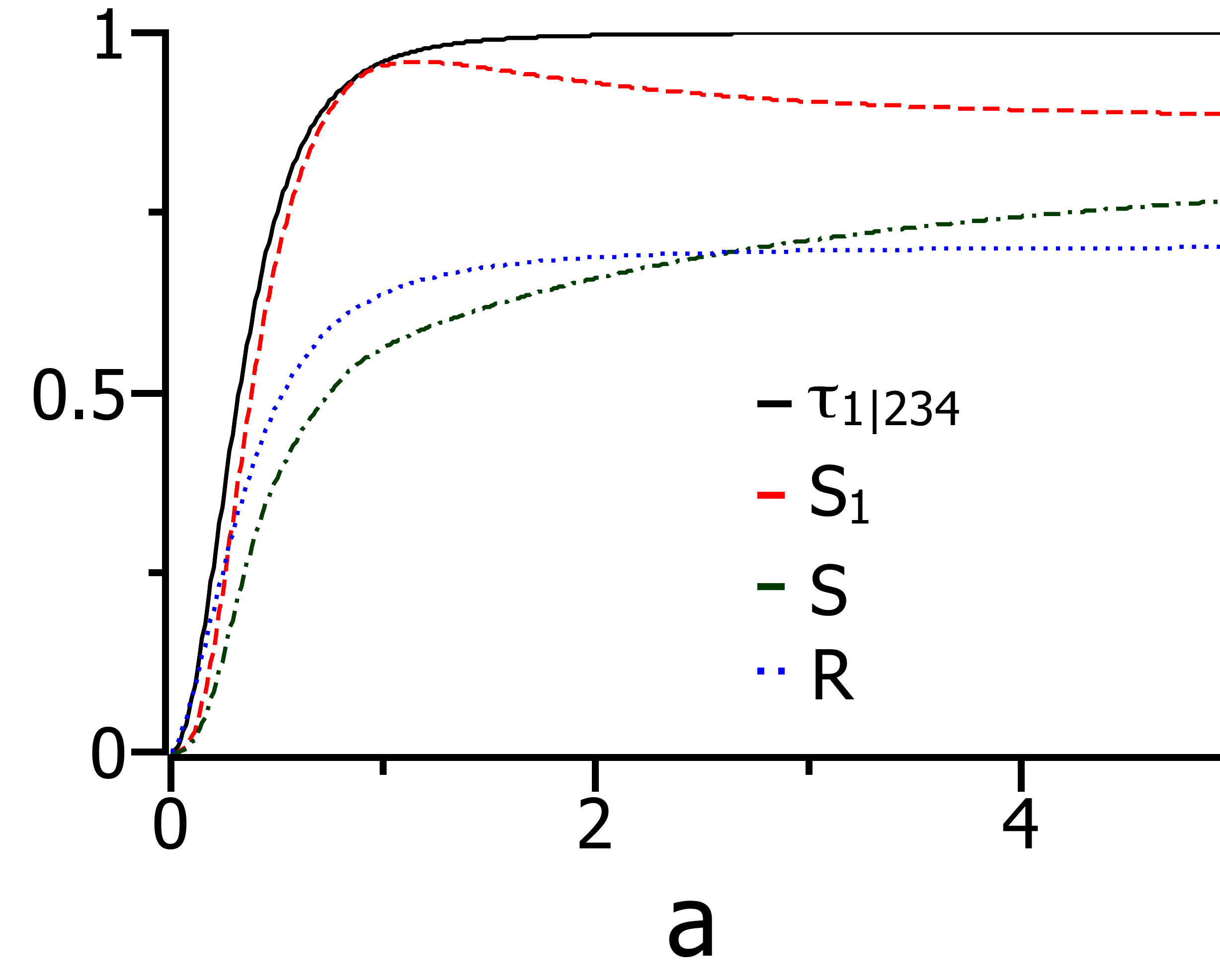}\caption{Plot of
one-tangle $\tau_{1|234}$ (black solid line), $S_{1}=\tau_{1|234}-\sum
_{j=2}^{4}\tau_{1|j}^{2}\left(  \rho_{1j}\right)  $ (Red dash), an estimate of
residual correlations S$=S_{1}-\left[  \frac{1}{2}\left(  \tau_{1|2|3}%
^{2}+\tau_{1|2|4}^{2}+\tau_{1|3|4}^{2}\right)  \right]  ^{\frac{1}{2}}$(Dark
Green Dash Dot), and partial four-way correlations R $=\sum_{j=2}^{4}%
\delta_{1j}$ (blue dot-dot), versus state parameter $a$ for the states
$L_{a,ia.\left(  ia\right)  _{2}}$.}%
\end{figure}

For these states, one tangle $\tau_{1|234}=\frac{8a^{2}+16a^{4}}{\left(
4a^{2}+1\right)  ^{2}}$ satisfies Eq. (\ref{1tann4}). All three tangles have
the same value that is $\tau_{1|2|3}=\tau_{1|3|4}=\tau_{1|2|4}=\frac{8a^{3}%
}{\left(  4a^{2}+1\right)  ^{2}}$ , genuine four-tangle $\tau_{1|2|3|4}%
^{\left(  1\right)  }=0$ and $\left(  \tau_{1|2|3|4}^{\left(  0\right)
}\right)  ^{2}=\frac{4a^{4}}{\left(  4a^{2}+1\right)  ^{2}}$. One can verify
that (Eq. (\ref{n81jc}))
\[
4n_{8}\left(  \rho_{12}\right)  =\frac{1}{4}\tau_{1|2|3}^{2}+\frac{1}{4}%
\tau_{1|2|4}^{2}+\frac{1}{8}\left(  \tau_{1|2|3|4}^{\left(  2\right)  }\left(
\rho_{12}\right)  \right)  ^{2},
\]
where $\tau_{1|2|3|4}^{\left(  2\right)  }\left(  \rho_{12}\right)
=\frac{8\sqrt{3}a^{4}}{\left(  4a^{2}+1\right)  ^{2}}$ and $\tau
_{1|2|3|4}^{\left(  3\right)  }\left(  \rho_{12}\right)  =0$. However
$\tau_{1|2|3|4}^{\left(  3\right)  }\left(  \rho_{13}\right)  =\tau
_{1|2|3|4}^{\left(  3\right)  }\left(  \rho_{14}\right)  =\frac{4\sqrt{5}%
a^{4}}{\left(  4a^{2}+1\right)  ^{2}}$, while the value of four-tangle
$\tau_{1|2|3|4}^{\left(  2\right)  }\left(  \rho_{13}\right)  =\tau
_{1|2|3|4}^{\left(  2\right)  }\left(  \rho_{14}\right)  =\frac{4a^{3}%
\sqrt{6a^{2}+10}}{\left(  4a^{2}+1\right)  ^{2}}$ consistent with
$4n_{8}\left(  \rho_{13}\right)  =4n_{8}\left(  \rho_{14}\right)  $. By using
the values of $n_{4}\left(  \rho_{1j}\right)  $ calculated from reduced
two-qubit states,%
\[
n_{4}\left(  \rho_{12}\right)  =\frac{4\left(  a^{4}+a^{2}\right)  }{\left(
4a^{2}+1\right)  ^{2}};n_{4}\left(  \rho_{13}\right)  =n_{4}\left(  \rho
_{14}\right)  =\frac{\left(  2a^{2}+7a^{4}\right)  }{\left(  4a^{2}+1\right)
^{2}},
\]
net four-way correlations can be calculated from the constraint (Eq.
(\ref{sumn4c})),%
\[
\sum_{j=2}^{4}\left(  n_{4}\left(  \rho_{1j}\right)  -\tau_{1|j}^{2}\left(
\rho_{1j}\right)  \right)  ^{2}-\frac{1}{2}\left(  \tau_{1|2|3}^{2}%
+\tau_{1|2|4}^{2}+\tau_{1|3|4}^{2}\right)  =\sum_{j=2}^{4}\Delta_{1j}.
\]
Figure (1) displays one-tangle $\tau_{1|234}$, the sum of three-way and
four-way correlations $S_{1}=\tau_{1|234}-\sum_{j=2}^{4}\tau_{1|j}^{2}\left(
\rho_{1j}\right)  $ (Eq. (\ref{S1})), estimated four-way correlations
$S=S_{1}-\left[  \frac{1}{2}\left(  \tau_{1|2|3}^{2}+\tau_{1|2|4}^{2}%
+\tau_{1|3|4}^{2}\right)  \right]  ^{\frac{1}{2}}$, and partial residual
four-way correlations quantified by $R={\left[  \sum_{j=2}^{4}\delta
_{1j}\right]  }^{\frac{1}{2}}$, versus state parameter $a$ for the states
$L_{a,ia.\left(  ia\right)  _{2}}$. We notice that $S\geq0$ and $R\geq0$ for
all the values of $a$, as expected.

\section{Classification of Four-qubit States}

\begin{table}[ptb]
\caption{ Entanglement measures for four-qubit entangled states. Class refers
to group of states \cite{vers02} which gives rise to tangles when qubit
A$_{1}$ is the focus qubit. Qubit A$_{1}$ is separable in class
$L_{0_{3+\overline{1}}0_{3+\overline{1}}}$.}%
\centering
\par%
\begin{tabular}
[c]{||l||l||l||l||l||l||l||l||l||}\hline\hline
$%
\begin{array}
[c]{c}%
\text{Class}\Rightarrow\\
\text{Tangle}\Downarrow
\end{array}
$ & $L_{abc_{2}}$ & $L_{ab_{3}}$ & $L_{a_{4}}$ & $L_{0_{7\oplus1}}$ &
$L_{0_{5\oplus3}}$ & $L_{a_{2}b_{2}}$ & $G_{abcd}$ & $L_{a_{2}0_{3\oplus1}}%
$\\\hline\hline
$\tau_{1|2|3|4}^{\left(  2\right)  }\left(  \rho_{1j}\right)  $ & $\neq0$ &
$\neq0$ & $\neq0$ & $\neq0$ & $\neq0$ & $\neq0$ & $\neq0$ & $\neq
0$\\\hline\hline
$\tau_{1|2|3|4}^{\left(  1\right)  }$ & $\neq0$ & \textbf{0} & \textbf{0} &
\textbf{0} & \textbf{0} & $\neq0$ & $\neq0$ & $\neq0$\\\hline\hline
$\tau_{1|2|3|4}^{\left(  3\right)  }\left(  \rho_{13}\right)  $ & $\neq0$ &
$\neq0$ & \textbf{0} & \textbf{0} & \textbf{0} & $\neq0$ & $\neq0$ & $\neq
0$\\\hline\hline
$\tau_{1|2|3|4}^{\left(  3\right)  }\left(  \rho_{12}\right)  ,\tau
_{1|2|3|4}^{\left(  3\right)  }\left(  \rho_{14}\right)  $ & $\neq0$ & $\neq0$
& $\neq0$ & \textbf{0} & \textbf{0} & $\neq0$ & $\neq0$ & $\neq0$%
\\\hline\hline
$\tau_{1|3|4}$ & $\neq0$ & $\neq0$ & $\neq0$ & $\neq0$ & \textbf{0} &
\textbf{0} & \textbf{0} & \textbf{0}\\\hline\hline
$\tau_{1|2|3}$ & $\neq0$ & $\neq0$ & $\neq0$ & $\neq0$ & $\neq0$ & \textbf{0}
& \textbf{0} & \textbf{0}\\\hline\hline
$\tau_{1|2|4}$ & $\neq0$ & $\neq0$ & $\neq0$ & $\neq0$ & $\neq0$ & $\neq0$ &
\textbf{0} & \textbf{0}\\\hline\hline
$\tau_{1|2}^{2}$ & $\neq0$ & $\neq0$ & $\neq0$ & \textbf{0} & \textbf{0} &
$\neq0$ & $\neq0$ & $\neq0$\\\hline\hline
$\tau_{1|3}^{2}$ & $\neq0$ & $\neq0$ & $\neq0$ & \textbf{0} & $\neq0$ &
$\neq0$ & $\neq0$ & $\neq0$\\\hline\hline
$\tau_{1|2}^{2}$ & $\neq0$ & $\neq0$ & $\neq0$ & \textbf{0} & $\neq0$ &
$\neq0$ & $\neq0$ & $\neq0$\\\hline\hline
$n_{16}\left(  \rho_{12}\right)  $ & $\neq0$ & $\neq0$ & $\neq0$ & \textbf{0}
& \textbf{0} & $\neq0$ & $\neq0$ & \textbf{0}\\\hline\hline
$n_{16}\left(  \rho_{13}\right)  $ & $\neq0$ & $\neq0$ & \textbf{0} &
\textbf{0} & \textbf{0} & $\neq0$ & $\neq0$ & \textbf{0}\\\hline\hline
$n_{16}\left(  \rho_{14}\right)  $ & $\neq0$ & $\neq0$ & $\neq0$ & \textbf{0}
& \textbf{0} & $\neq0$ & $\neq0$ & \textbf{0}\\\hline\hline
\end{tabular}
\end{table}Three-qubit states have been shown to belong to six equivalent
classes under stochastic local operations and classical communication (SLOCC)
In \cite{dur00}. However, $N>3$, there are infinite SLOCC classes \cite{li13}
, as such it is highly desirable to partition the infinite classes into a
finite number of families. \cite{vers02}, Verstraete et al. have shown that
there are nine families in four-qubit entanglement.\ Residue $\delta_{1j}$ is
greater or equal to weighted sum of four tangles. These tangles are natural
labels for fully entangled four-qubit states. Besides that, three tangles, and
two-tangles of sub-systems are helpful to understand the extent to which the
states can be manipulated by local operations. In a recent articles
\cite{ghah16} a classification of four-qubit states based on graph states is
given. The underlying spirit of this classification is similar to ours,
however our classification has the advantage that the entanglement quantifiers
obey the monogamy constraints listed in Eqs. (\ref{n81jc},\ref{sum4n8c}%
,\ref{n41jc},\ref{sumn4c},\ref{1tanc1},\ref{1tanc2}).

We notice that the four qubit states may be grouped together in finite number
of entanglement types by using a set of unitary invariant functions of state
coefficients to label the states. The set contains - two-tangles to quantify
the entanglement of two qubit sub-systems, $\tau_{i|j|k}$ to quantify
entanglement of sub-systems due to three-way correlations, and $\tau
_{1|2|3|4}^{\left(  1\right)  }$, $\tau_{1|2|3|4}^{(2)}\left(  \rho
_{1j}\right)  $, and $\tau_{1|2|3|4}^{\left(  3\right)  }\left(  \rho
_{1j}\right)  $ ($j=2-4$) to quantify entanglement due to four-way
correlations. Besides that we can also use the unitary invariant coefficients
$n_{12}\left(  \rho_{1j}\right)  $ and $n_{16}\left(  \rho_{1j}\right)  $,
$\quad j=2,3,4$, to distinguish between different entanglement types. On the
basis of these invariants, four-qubit entangled states lie in the following groups:

\begin{itemize}
\item[Group I] - Two-qubit and three qubit subsystems are entangled. Some or
all of the four-tangles are non-zero.

\item[Group II] - Three qubit subsystems are entangled. Some or all of the
four-tangles are non-zero.

\item[Group III] - Two-qubit subsystems are entangled. Some or all of the
four-tangles are non-zero.

\item[Group IV] - Two-qubit subsystems are entangled. All of the four-tangles
are zero. Four-way entanglement arises due to pair-wise entanglement.
\end{itemize}

The states in the nine classes of four-qubit states \cite{vers02} fit in
Groups I, II and III. Table II is a list of tangles useful to label eight
classes of four-qubit entangled states \cite{vers02}. The parameters
characterizing the states are taken to have distinct values. With the
exception of $L_{0_{3+\overline{1}}0_{3+\overline{1}}}$ states, the
four-tangle $\tau_{1|2|3|4}^{\left(  2\right)  }\left(  \rho_{1j}\right)
\neq0$ on nine classes of states listed in ref. \cite{vers02}. Class
$L_{abc_{2}}$ contains states on which all four tangles, three tangles, and
two-tangles are non-zero. States in classes $L_{ab_{3}}$, $L_{a_{4}}$,
$L_{0_{7\oplus1}}$, and $L_{0_{5\oplus3}}$ have $\tau_{1|2|3|4}^{\left(
1\right)  }\left(  \rho_{1j}\right)  =0$. State $L_{0_{5\oplus3}}$ is special
in that the four-tangle $\tau_{1|2|3|4}^{\left(  2\right)  }\left(  \rho
_{1j}\right)  $ is product of three tangles. For states in $G_{abcd}$ and
$L_{a_{2}0_{3\oplus1}}$, three-qubit subsystems have zero three tangles. An
example of states in Group IV is maximally entangled W-state,%
\[
\left\vert \widetilde{W}\right\rangle =\frac{1}{2}\left(  \left\vert
0000\right\rangle +\left\vert 1100\right\rangle +\left\vert 1010\right\rangle
+\left\vert 1001\right\rangle \right)  .
\]

All four-tangles and three-tangles are zero on state $\left\vert \widetilde
{W}\right\rangle $, and four-way entanglement is due to two-way correlations.
The state $L_{0_{3+\overline{1}}0_{3+\overline{1}}}$ does not have four-qubit entanglement.

\section{Entanglement transfer to environment}

\begin{figure}[th]
\centering
\includegraphics[height=4.0in]{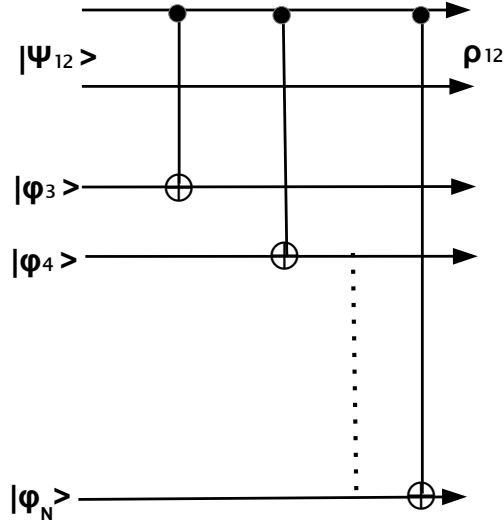}\caption{Circuit model for
entanglement transfer}%
\label{fig2}%
\end{figure}

\begin{figure}[th]
\centering
\includegraphics[height=3.5in]{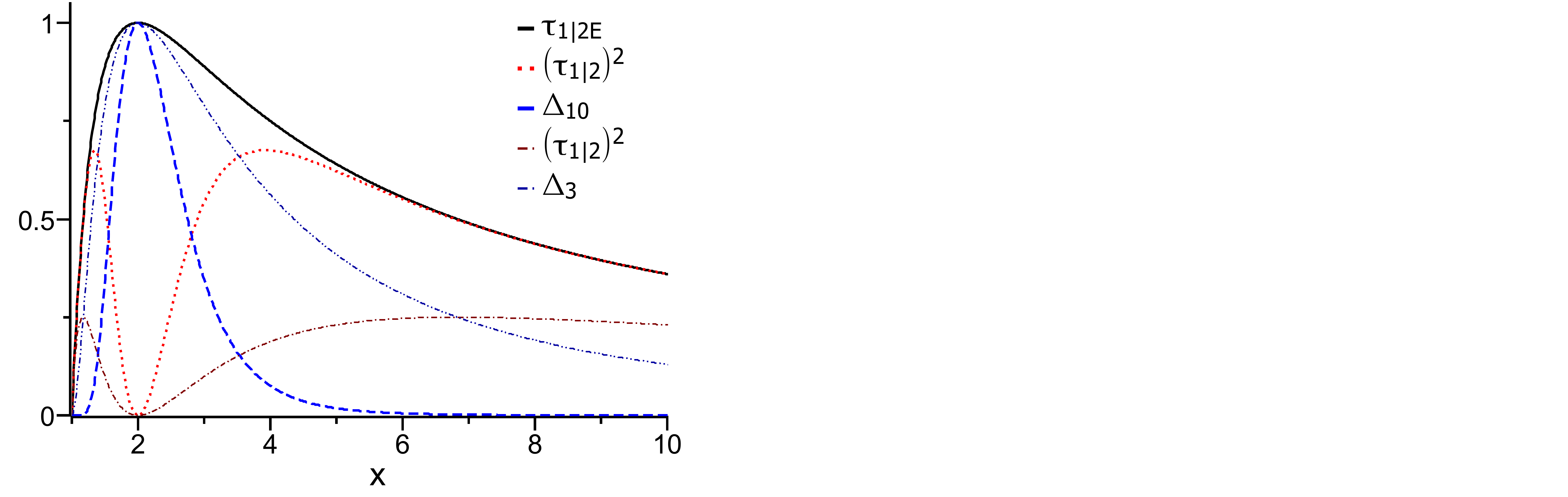} \caption{Plot of one-tangle
$\tau_{1|234}$ (black solid line), $\tau_{1|2}^{2}$ (red dash), Residual
correlations $\Delta$ (blue dots) for $M=8$ and $\tau_{1|2}^{2}$ (wine dash
dot), Residual correlations $\Delta$ (navy blue dash dot dot) for $M=1$.}%
\label{fig3}%
\end{figure}

If one of the two entangled qubits, interacts successively with environment
qubits resulting in an increase in residual correlations, then the
entanglement of the pair tends to zero. Here we present a toy model that uses
CNOT Gate to generate additional correlations between qubit one and the
environment represented by a product state of additional qubits. For
simplicity a single parameter model with two qubits, initially in pure state
$\left\vert \Psi_{12}\right\rangle =\frac{1}{\sqrt{x}}\left(  \left\vert
00\right\rangle +\sqrt{x-1}\left\vert 11\right\rangle \right)  $, is
considered. Environment qubits are in a product state $%
{\displaystyle\prod\limits_{j}}
$ $\left\vert \phi_{j}\right\rangle $ where $\left\vert \phi_{j}\right\rangle
=\frac{1}{\sqrt{x}}\left(  \left\vert 0\right\rangle +\sqrt{x-1}\left\vert
1\right\rangle \right)  $, as such the initial state of the system%
\[
\left\vert \Psi_{12}\right\rangle \left\vert E\right\rangle =\frac{1}{\sqrt
{x}}\left(  \left\vert 00\right\rangle +\sqrt{x-1}\left\vert 11\right\rangle
\right)  \left\vert \phi_{3}\right\rangle \left\vert \phi_{4}\right\rangle
...\left\vert \phi_{N}\right\rangle ,
\]
is an entangled state with one tangle $\tau_{1|2E}=\tau_{1|2}^{2}$ given by
\[
\tau_{1|2}^{2}=\frac{4\left(  x-1\right)  }{x^{2}}=n_{4}\left(  \rho
_{12}\right)  ;\quad n_{8}\left(  \rho_{12}\right)  =0;x\neq0,1.
\]
A CNOT gate on qubit pair $A_{1}A_{3}$ generates three-way correlations with a
decrease in two-tangle, while the value of one-tangle is not changed. After
this step the system is in state $\left\vert \Psi_{123}\right\rangle
\left\vert E_{N-1}\right\rangle $ and $\tau_{1|2}^{2}=\left(  \frac{4\left(
x-1\right)  }{x^{2}}\right)  ^{2}$. Next step is applying a CNOT to qubit pair
$A_{1}A_{4}$ with $A_{4}$ as target qubit. Successive applications of CNOT,
always with qubit $A_{1}$ as control qubit and one of the environment qubits
as target qubit, do not change $\tau_{1|2E}$, but generate correlations
distributed over a larger number of qubits. One can verify that after $M$
applications of CNOT ($M$ varies from $1$ to $N-2$), two tangles of the state
satisfy%
\[
\tau_{1|2}^{2}\left(  \rho_{12}\right)  =\left(  \frac{4\left(  x-1\right)
}{x^{2}}\right)  ^{M+1};\quad\tau_{1|j}^{2}\left(  \rho_{1j}\right)  =0\text{
for }j=3\text{ to N.}%
\]
whereas residual correlations are given by%
\[
\tau_{1|2E}-\tau_{1|2}^{2}\left(  \rho_{12}\right)  =\frac{4\left(
x-1\right)  }{x^{2}}-\left(  \frac{4\left(  x-1\right)  }{x^{2}}\right)
^{M+1}\text{.}%
\]
$\allowbreak\allowbreak$ $\allowbreak$ Fig. (3) displays $\tau_{1|2E}$,
$\tau_{1|2}^{2}\left(  \rho_{12}\right)  $ and residual correlations $\Delta$
for $M=1$ and $M=8$ as a function of variable $x$. One may notice that for
$x=2$, no entanglement transfer to additional qubits occurs. After eight steps
most of the two-way correlations have leaked to environment for $x=6$.
Notably, interaction with a single qubit ($M=1$) which generates only
three-way correlations, results in $\tau_{1|2}^{2}\left(  \rho_{12}\right)
=\Delta$ for $x_{1}=1.1716$ and $x_{2}=6.8284$.

\section{Concluding Remarks}

Two-tangle (refs. \cite{hill97,woot98}) of a two-qubit mixed state $\rho$ is a
known function of eigenvalues of non Hermitian matrix $\rho\widetilde{\rho}$
where $\widetilde{\rho}=\left(  \sigma_{y}\otimes\sigma_{y}\right)  \rho
^{\ast}\left(  \sigma_{y}\otimes\sigma_{y}\right)  $. As shown in ref.
\cite{arxiv2021} if the state $\rho$ is known to be part of an $N-$qubit
system in a pure state then the non-Hermitian matrix $\rho\widetilde{\rho}$
can be used to extract information about the correlations of the pair of
qubits with $\left(  N-2\right)  $ qubits in the state $\left\vert \Psi
_{N}\right\rangle $. In this article it is shown that if a two-qubit state is
a marginal state of a four-qubit pure state then the residual correlations are
quantified by well defined unitary invariant functions of state coefficients.

Our main result is the set of constraints on one-tangle of a focus qubit,
two-tangles, three-tangles and four-way correlations, obtained by expressing
the coefficients in the characteristic polynomial of $\rho\widetilde{\rho}$ in
terms of state coefficients of a four-qubit pure state. The tangles
characterizing a four-qubit state satisfy the constraints represented by Eqs.
(\ref{n81jc}, \ref{sum4n8c}, \ref{n41jc}, \ref{sumn4c}, \ref{1tanc1}) and Eq.
(\ref{1tanc2}). The residual four-qubit correlations obtained by subtracting
two-tangles and three-tangles as in Eqs. (\ref{n41jc}) represent contributions
from all possible four-qubit entanglement modes. One tangle of a four-qubit
pure state satisfies the constraints given by Eqs. (\ref{1tanc1}) and
(\ref{1tanc2}) independent of the class to which a given four-qubit state
belongs. In particular, these constraints are satisfied by the set of states
$L_{a,ia.\left(  ia\right)  _{2}}$ of ref. \cite{vers02} that violate the
entanglement monogamy relation obtained by generalizing the CKW inequality.
The difference between one tangle and contributions from two-tangles and
three-tangles in Eq. (\ref{1tanc2}) represents the residual correlations
beyond three-way correlations present in a 4-qubit pure state.

Four tangles $\tau_{1|2|3|4}^{(0)}$, $\tau_{1|2|3|4}^{(1)}$, $\tau
_{1|2|3|4}^{(2)}\left(  \rho_{1j}\right)  $ and $\tau_{1|2|3|4}^{(3)}\left(
\rho_{1j}\right)  $, ($j=2-4$), may be used to identify and label entangled
states that are equivalent under local unitary transformations. Local unitary
equivalence is an important marker to group together states with similar
properties. Using the elements of the set containing two-tangles,
three-tangles and four-tangles to label four-qubit states the states in nine
classes of four-qubit states \cite{vers02}, and W state are grouped together
in four-groups as shown in Table II. Using a simple circuit model, monogamy of
entanglement is also shown to result in loss of entanglement of a pair of
qubits when one of the qubits interacts successively with environment qubits.

This work reveals constraints on the sharing of entanglement at multiple
levels and offers insight into quantification of those features of quantum
correlations, which only emerge beyond the bipartite scenario. It will be
interesting to investigate the interplay between the entanglement trade-off
and frustration phenomena in complex quantum systems \cite{giam11}. Our
approach also paves the way to understanding scaling of entanglement
distribution as qubits are added to obtain larger multiqubit quantum systems.

\appendix{}

\section{Derivation of Eq. (\ref{n4rho})\label{A}}

To obtain Eq. (\ref{n4rho}) we use the expressions for the coefficients
$n_{d}\left(  \rho\right)  $ ($d=4,8,12,16$) in terms of eigenvalues of matrix
$\left(  \rho\widetilde{\rho}\right)  ,$ Eqs. (\ref{n4n8}), (\ref{n12}) and
(\ref{n16}) along with the condition $\lambda_{1}\geq\lambda_{2}\geq
\lambda_{3}\geq\lambda_{4}$ and $C\left(  \rho\right)  =\sqrt{\lambda_{1}%
}-\sqrt{\lambda_{2}}-\sqrt{\lambda_{3}}-\sqrt{\lambda_{4}}$, that is%

\begin{align}
&  n_{4}\left(  \rho\right)  -\left\vert C\left(  \rho\right)  \right\vert
^{2}=\left(  \lambda_{1}+\lambda_{2}+\lambda_{3}+\lambda_{4}\right)  -\left(
\sqrt{\lambda_{1}}-\sqrt{\lambda_{2}}-\sqrt{\lambda_{3}}-\sqrt{\lambda_{4}%
}\right)  ^{2}\nonumber\\
&  =2\sqrt{\left(
\begin{array}
[c]{c}%
\sqrt{\lambda_{1}}\left(  \sqrt{\lambda_{2}}+\sqrt{\lambda_{3}}+\sqrt
{\lambda_{4}}\right) \\
-\allowbreak\left(  \sqrt{\lambda_{2}}\sqrt{\lambda_{3}}+\sqrt{\lambda_{2}%
}\sqrt{\lambda_{4}}+\sqrt{\lambda_{3}}\sqrt{\lambda_{4}}\right)
\end{array}
\right)  ^{2}}\nonumber\\
&  =2\sqrt{n_{8}\left(  \rho\right)  +2\sqrt{n_{16}\left(  \rho\right)
}+2C\left(  \rho\right)  \left(
\begin{array}
[c]{c}%
\sqrt{\lambda_{1}}\sqrt{\lambda_{2}}\sqrt{\lambda_{3}}+\sqrt{\lambda_{1}}%
\sqrt{\lambda_{2}}\sqrt{\lambda_{4}}\\
+\sqrt{\lambda_{1}}\sqrt{\lambda_{3}}\sqrt{\lambda_{4}}-\allowbreak
\sqrt{\lambda_{2}}\sqrt{\lambda_{3}}\sqrt{\lambda_{4}}%
\end{array}
\right)  }%
\end{align}
Furthermore, we can rewrite the expression given above as%
\begin{equation}
n_{4}\left(  \rho\right)  =\left\vert C\left(  \rho\right)  \right\vert
^{2}+2\sqrt{%
\begin{array}
[c]{c}%
n_{8}\left(  \rho\right)  +2\sqrt{n_{16}\left(  \rho\right)  }\\
+2C\left(  \rho\right)  \sqrt{n_{12}\left(  \rho\right)  +\sqrt{n_{16}\left(
\rho\right)  }\left(  n_{4}\left(  \rho\right)  -\left\vert C\left(
\rho\right)  \right\vert ^{2}\right)  }%
\end{array}
.} \label{n4ro}%
\end{equation}
Substituting $C\left(  \rho\right)  =\pm\left\vert C\left(  \rho\right)
\right\vert $ in Eq. (\ref{n4ro}), the coefficient $n_{4}\left(  \rho\right)
$ satisfies the relation%

\begin{equation}
n_{4}\left(  \rho\right)  =\left\vert C\left(  \rho\right)  \right\vert
^{2}+2\sqrt{n_{8}\left(  \rho\right)  +2\sqrt{n_{16}\left(  \rho\right)  }%
\pm2\sqrt{f_{16}\left(  \rho\right)  }},
\end{equation}
where%
\begin{equation}
f_{16}\left(  \rho\right)  =\left\vert C\left(  \rho\right)  \right\vert
^{2}\left(  n_{12}\left(  \rho\right)  +\sqrt{n_{16}\left(  \rho\right)
}\left(  n_{4}\left(  \rho\right)  -\left\vert C\left(  \rho\right)
\right\vert ^{2}\right)  \right)  .
\end{equation}

\section{Expressions for $n_{4}\left(  \rho_{1j}\right)  $, $n_{8}\left(
\rho_{1j}\right)  $, $\left\{  I_{A_{4}}^{4-m,m}:m=0,4\right\}  $,
$N_{4,8}^{\left(  1jk\right)  }$ and $P_{1j}$ in terms of two-qubit unitary
invariants \label{B}}

\subsection{Notation\label{1}}

In this subsection we set up the notation used to express the relevant three
and four-qubit invariants in terms of two-qubit unitary invariants. In section
(\ref{fourstate}) a general four-qubit pure state was written as%
\begin{equation}
\left\vert \Psi_{1234}\right\rangle =\sum_{i_{1},i_{2},i_{3}}\left(
a_{i_{1}i_{2}i_{3}0}\left\vert i_{1}i_{2}i_{3}0\right\rangle +a_{i_{1}%
i_{2}i_{3}1}\left\vert i_{1}i_{2}i_{3}1\right\rangle \right)  ,\quad\left(
i_{m}=0,1\right)  ,
\end{equation}
and the determinants of negativity fonts of the state defined as $D_{\left(
A_{3}\right)  _{i_{3}}\left(  A_{4}\right)  _{i_{4}}}^{00}=a_{00i_{3}i_{4}%
}a_{11i_{3}i_{4}}-a_{10i_{3}i_{4}}a_{01i_{3}i_{4}}$ (two-way), $D_{\left(
A_{2}\right)  _{i_{2}}\left(  A_{4}\right)  _{i_{4}}}^{00}=a_{0i_{2}0i_{4}%
}a_{1i_{2}1i_{4}}-a_{1i_{2}0i_{4}}a_{0i_{2}1i_{4}}$ (two-way), $D_{\left(
A_{2}\right)  _{i_{2}}\left(  A_{3}\right)  _{i_{3}}}^{00}=a_{0i_{2}i_{3}%
0}a_{1i_{2}i_{3}1}-a_{1i_{2}i_{3}0}a_{0i_{2}i_{3}1}$ (two-way), $D_{\left(
A_{4}\right)  _{i_{4}}}^{00i_{3}}=a_{00i_{3}i_{4}}a_{11,i_{3}\oplus1,i_{4}%
}-a_{10i_{3}i_{4}}a_{01,i_{3}\oplus1,i_{4}}$ (three-way), $D_{\left(
A_{3}\right)  _{i_{3}}}^{00i_{4}}=a_{00i_{3}i_{4}}a_{11i_{3},i_{4}\oplus
1}-a_{10i_{3}i_{4}}a_{01i_{3},i_{4}\oplus1}$ (three-way), $D_{\left(
A_{2}\right)  _{i_{2}}}^{00i_{4}}=a_{0i_{2}0i_{4}}a_{1i_{2}1i_{4}\oplus
1}-a_{1i_{2}0i_{4}}a_{0i_{2}1i_{4}\oplus1}$ (three-way), and $D^{00i_{3}i_{4}%
}=a_{00i_{3}i_{4}}a_{11,i_{3}\oplus1,i_{4}\oplus1}-a_{10i_{3}i_{4}}%
a_{01,i_{3}\oplus1,i_{4}\oplus1}$- (four-way). The notation for two-qubit
unitary invariants for qubit pairs $A_{1}A_{2}$, $A_{1}A_{3}$, and $A_{1}%
A_{4}$ follows. The set of invariants with respect to unitary transformations
on qubits $A_{1}$ and $A_{2}$ are given by%
\begin{equation}
E_{2}=D_{\left(  A_{3}\right)  _{0}\left(  A_{4}\right)  _{0}}^{00}%
,D_{2}=D_{\left(  A_{3}\right)  _{1}\left(  A_{4}\right)  _{1}}^{00}%
,C_{2}=D_{\left(  A_{3}\right)  _{1}\left(  A_{4}\right)  _{0}}^{00}%
,B_{2}=D_{\left(  A_{3}\right)  _{0}\left(  A_{4}\right)  _{1}}^{00},
\end{equation}%
\begin{equation}
F_{2}=D_{\left(  A_{4}\right)  _{0}}^{000}+D_{\left(  A_{4}\right)  _{0}%
}^{001},L_{2}=D_{\left(  A_{4}\right)  _{1}}^{000}+D_{\left(  A_{4}\right)
_{1}}^{001},
\end{equation}%
\begin{equation}
G_{2}=D_{\left(  A_{3}\right)  _{0}}^{000}+D_{\left(  A_{3}\right)  _{0}%
}^{001},K_{2}=D_{\left(  A_{3}\right)  _{1}}^{000}+D_{\left(  A_{3}\right)
_{1}}^{001},
\end{equation}%
\begin{equation}
H_{02}=D^{0000}+D^{0011},H_{12}=D^{0001}+D^{0010}.
\end{equation}
Two-qubit invariants with respect to unitaries on qubits A$_{1}$ and A$_{3}$
are denoted by%
\begin{equation}
E_{3}=D_{\left(  A_{2}\right)  _{0}\left(  A_{4}\right)  _{0}}^{00}%
,D_{3}=D_{\left(  A_{2}\right)  _{1}\left(  A_{4}\right)  _{1}}^{00}%
,C_{3}=D_{\left(  A_{2}\right)  _{1}\left(  A_{4}\right)  _{0}}^{00}%
,B_{3}=D_{\left(  A_{2}\right)  _{0}\left(  A_{4}\right)  _{1}}^{00},
\end{equation}%
\begin{equation}
F_{3}=D_{\left(  A_{4}\right)  _{0}}^{000}-D_{\left(  A_{4}\right)  _{0}%
}^{001},L_{3}=D_{\left(  A_{4}\right)  _{1}}^{000}-D_{\left(  A_{4}\right)
_{1}}^{001},
\end{equation}%
\begin{equation}
G_{3}=D_{\left(  A_{2}\right)  _{0}}^{000}+D_{\left(  A_{2}\right)  _{0}%
}^{001},K_{3}=D_{\left(  A_{2}\right)  _{1}}^{000}+D_{\left(  A_{4}\right)
_{1}}^{001},
\end{equation}%
\begin{equation}
H_{03}=D^{0000}-D^{0010},H_{13}=D^{0001}-D^{0011}.
\end{equation}
Invariants with respect to unitaries on qubits A$_{1}$ and A$_{4}$ read as%
\begin{equation}
E_{4}=D_{\left(  A_{2}\right)  _{0}\left(  A_{3}\right)  _{0}}^{00}%
,D_{4}=D_{\left(  A_{2}\right)  _{1}\left(  A_{3}\right)  _{1}}^{00}%
,C_{4}=D_{\left(  A_{2}\right)  _{1}\left(  A_{3}\right)  _{0}}^{00}%
,B_{4}=D_{\left(  A_{2}\right)  _{0}\left(  A_{3}\right)  _{1}}^{00},
\end{equation}%
\begin{equation}
F_{4}=D_{\left(  A_{3}\right)  _{0}}^{000}-D_{\left(  A_{3}\right)  _{0}%
}^{001},L_{4}=D_{\left(  A_{3}\right)  _{1}}^{000}-D_{\left(  A_{3}\right)
_{1}}^{001},
\end{equation}%
\begin{equation}
G_{4}=D_{\left(  A_{2}\right)  _{0}}^{000}-D_{\left(  A_{2}\right)  _{0}%
}^{001},K_{4}=D_{\left(  A_{2}\right)  _{1}}^{000}-D_{\left(  A_{4}\right)
_{1}}^{001},
\end{equation}%
\begin{equation}
H_{04}=D^{0000}-D^{0001},H_{14}=D^{0010}-D^{0011},
\end{equation}

\subsection{The coefficients $n_{4}\left(  \rho_{1j}\right)  $ and
$n_{8}\left(  \rho_{1j}\right)  $\label{2}}

The coefficient $n_{4}\left(  \rho_{1j}\right)  =tr\left(  \rho_{1j}%
\widetilde{\rho_{1j}}\right)  $ is found to have the form%

\begin{align}
n_{4}\left(  \rho_{1j}\right)   &  =4\left(  \left\vert E_{j}\right\vert
^{2}+\left\vert B_{j}\right\vert ^{2}+\left\vert C_{j}\right\vert
^{2}+\left\vert D_{j}\right\vert ^{2}\right)  +2\left(  \left\vert
G_{j}\right\vert ^{2}+\left\vert K_{j}\right\vert ^{2}\right) \nonumber\\
&  +2\left(  \left\vert F_{j}\right\vert ^{2}+\left\vert L_{j}\right\vert
^{2}\right)  +\left\vert H_{0j}+H_{1j}\right\vert ^{2}+\left\vert
H_{0j}-H_{1j}\right\vert ^{2}. \label{n4form}%
\end{align}
Maximum value of $n_{4}\left(  \rho_{1j}\right)  $ is one.

The degree eight coefficient $n_{8}\left(  \rho_{1j}\right)  =\frac{1}%
{2}\left(  tr\left(  \rho_{1j}\widetilde{\rho_{1j}}\right)  \right)
^{2}-\frac{1}{2}tr\left(  \left(  \rho_{1j}\widetilde{\rho_{1j}}\right)
^{2}\right)  $, which is a function of three-qubit invariants reads as%
\begin{align}
n_{8}\left(  \rho_{1j}\right)   &  =\left\vert G_{j}^{2}-4E_{j}B_{j}%
\right\vert ^{2}+\left\vert K_{j}^{2}-4C_{j}D_{j}\right\vert ^{2}+\left\vert
F_{j}^{2}-4E_{j}C_{j}\right\vert ^{2}\nonumber\\
&  +\left\vert L_{j}^{2}-4B_{j}D_{j}\right\vert ^{2}+\left\vert H_{0j}%
^{2}-4E_{j}D_{j}\right\vert ^{2}+\left\vert H_{1j}^{2}-4B_{j}C_{j}\right\vert
^{2}\nonumber\\
&  +2\left\vert G_{j}K_{j}-F_{j}L_{j}\right\vert ^{2}+2\left\vert H_{0j}%
H_{1j}-G_{j}K_{j}\right\vert ^{2}+2\left\vert \left(  H_{0j}H_{1j}-F_{j}%
L_{j}\right)  \right\vert ^{2}\nonumber\\
&  +2\left\vert F_{j}G_{j}-2E_{j}H_{1j}\right\vert ^{2}+2\left\vert F_{j}%
K_{j}-2C_{j}H_{0j}\right\vert ^{2}+2\left\vert G_{j}L_{j}-2B_{j}%
H_{0j}\right\vert ^{2}\nonumber\\
&  +2\left\vert K_{j}L_{j}-2H_{1j}D_{j}\right\vert ^{2}+2\left\vert
H_{0j}F_{j}-2E_{j}K_{j}\right\vert ^{2}+2\left\vert H_{0j}G_{j}-2E_{j}%
L_{j}\right\vert ^{2}\nonumber\\
&  +2\left\vert H_{0j}K_{j}-2F_{j}D_{j}\right\vert ^{2}+2\left\vert
H_{0j}L_{j}-2G_{j}D_{j}\right\vert ^{2}+2\left\vert H_{1j}F_{j}-2C_{j}%
G_{j}\right\vert ^{2}\nonumber\\
&  +2\left\vert H_{1j}K_{j}-2C_{j}L_{j}\right\vert ^{2}+2\left\vert
H_{1j}G_{j}-2B_{j}F_{j}\right\vert ^{2}+2\left\vert H_{1j}L_{j}-2B_{j}%
K_{j}\right\vert ^{2} \label{n8form}%
\end{align}
One can verify that $0\leq16n_{8}\left(  \rho_{1j}\right)  \leq1.$

\subsection{Degree four three-qubit invariants $\left\{  I_{A_{4}}%
^{4-m,m}:m=0,4\right\}  $\label{3}}

Degree four three-qubit invariants of a four-qubit state relevant to
constructing the upper bound on $\tau_{1|2|3}\left(  \rho_{123}\right)  $ in
terms of two-qubit invariants for the pair $A_{1}A_{2}$ are listed below:%

\begin{equation}
I_{A_{4}}^{4,0}=F_{2}^{2}-4E_{2}C_{2};I_{A_{4}}^{0,4}=L_{2}^{2}-4B_{2}D_{2},
\end{equation}%
\begin{equation}
I_{A_{4}}^{3,1}=\frac{1}{2}F_{2}\left(  H_{02}+H_{12}\right)  -\left(
E_{2}K_{2}+C_{2}G_{2}\right)  ,
\end{equation}%
\begin{equation}
I_{A_{4}}^{1,3}=\frac{1}{2}L_{2}\left(  H_{02}+H_{12}\right)  -\left(
B_{2}K_{2}+D_{2}G_{2}\right)  ,
\end{equation}
and%
\begin{equation}
I_{A_{4}}^{2,2}=\frac{1}{6}\left(  H_{02}+H_{12}\right)  ^{2}-\frac{2}{3}%
G_{2}K_{2}+\frac{1}{3}F_{2}L_{2}-\frac{2}{3}\left(  E_{2}D_{2}+B_{2}%
C_{2}\right)  .
\end{equation}

\subsection{Degree eight invariants $N_{4,8}^{\left(  123\right)  }%
,N_{4,8}^{\left(  134\right)  }$ and $N_{4,8}^{\left(  143\right)  }$%
\label{4}}

In order to write down the coefficients $n_{8}\left(  \rho_{1j}\right)
$,$\left\{  j=2-4\right\}  $, we need the form of $N_{4,8}^{\left(
123\right)  },N_{4,8}^{\left(  134\right)  }$ and $N_{4,8}^{\left(
143\right)  }$. The coefficients $N_{4,8}^{\left(  123\right)  }$ and
$N_{4,8}^{\left(  143\right)  }$ are obtained by substituting, respectively,
$j=2$ and $4$ in the following equation:
\begin{align}
N_{4,8}^{\left(  1j3\right)  }  &  =\left\vert \left(  F_{j}^{2}-4E_{j}%
C_{j}\right)  \right\vert ^{2}+\left\vert \left(  H_{0j}+H_{1j}\right)
F_{j}-2E_{j}K_{j}-2C_{j}G_{j}\right\vert ^{2}\nonumber\\
&  +\frac{1}{6}\left\vert \left(  H_{0j}+H_{1j}\right)  ^{2}-4G_{j}%
K_{j}+2F_{j}L_{j}-4B_{j}C_{j}-4E_{j}D_{j}\right\vert ^{2}\nonumber\\
&  +\left\vert \left(  H_{0j}+H_{1j}\right)  L_{j}-2G_{j}D_{j}-2B_{j}%
K_{j}\right\vert ^{2}+\left\vert L_{j}^{2}-4B_{j}D_{j}\right\vert ^{2},
\label{N481j3}%
\end{align}
whereas $N_{4,8}^{\left(  124\right)  }$ is given by%
\begin{align}
N_{4,8}^{\left(  124\right)  }  &  =\left\vert \left(  G_{3}^{2}-4E_{3}%
B_{3}\right)  \right\vert ^{2}+\left\vert \left(  H_{03}+H_{13}\right)
G_{3}-2E_{3}L_{3}-2B_{3}F_{3}\right\vert ^{2}\nonumber\\
&  +\frac{1}{6}\left\vert \left(  H_{03}+H_{13}\right)  ^{2}+2G_{3}%
K_{3}-4F_{3}L_{3}-4E_{3}D_{3}-4B_{3}C_{3}\right\vert ^{2}\nonumber\\
&  +\left\vert \left(  H_{03}+H_{13}\right)  K_{3}-2F_{3}D_{3}-2C_{3}%
L_{3}\right\vert ^{2}+\left\vert \left(  K_{3}^{2}-4C_{3}D_{3}\right)
\right\vert ^{2}. \label{N48124}%
\end{align}

\subsection{Invariants $P_{1j}\ $and $M_{4,8}\left(  \rho_{1j}\right)  $
\label{5}}

Invariants $P_{1j}$ are degree four functions of determinants of negativity
fonts and read as%
\begin{equation}
P_{1j}=\left(  H_{0j}+H_{1j}\right)  ^{2}-4F_{j}L_{j}-4G_{j}K_{j}+8E_{j}%
D_{j}+8B_{j}C_{j}.
\end{equation}

Term $M_{4,8}\left(  \rho_{1j}\right)  $ is a sum of three-qubit invariants,
that is%
\begin{align}
M_{4,8}\left(  \rho_{1j}\right)   &  =2\left\vert F_{j}G_{j}-2E_{j}%
H_{1j}\right\vert ^{2}+\left\vert \left(  \left(  H_{1j}-H_{0j}\right)
G_{j}+2E_{j}L_{j}-2B_{j}F_{j}\right)  \right\vert ^{2}\nonumber\\
&  +2\left\vert G_{j}L_{j}-2B_{j}H_{0j}\right\vert ^{2}+\left\vert \left(
\left(  H_{1j}-H_{0j}\right)  F_{j}+2E_{j}K_{j}-2C_{j}G_{j}\right)
\right\vert ^{2}\nonumber\\
&  +\frac{1}{2}\left\vert H_{1j}^{2}-H_{0j}^{2}+4E_{j}D_{j}-4B_{j}%
C_{j}\right\vert ^{2}\nonumber\\
&  +\left\vert \left(  \left(  H_{1j}-H_{0j}\right)  L_{j}+2G_{j}D_{j}%
-2B_{j}K_{j}\right)  \right\vert ^{2}+2\left\vert F_{j}K_{j}-2C_{j}%
H_{0j}\right\vert ^{2}\nonumber\\
&  +\left\vert \left(  H_{1j}-H_{0j}\right)  K_{j}+2F_{j}D_{j}-2C_{j}%
L_{j}\right\vert ^{2}+2\left\vert K_{j}L_{j}-2D_{j}H_{1j}\right\vert ^{2}.
\end{align}

\end{document}